\documentclass[twocolumn]{aastex631}       \newif\ifdraft \drafttrue
\usepackage{graphicx}
\usepackage{xcolor}
\usepackage{upgreek}
\usepackage{amsmath}	
\ifdraft
   \usepackage{hyperref}
   \hypersetup{
               colorlinks=true,
               linkcolor=black,
               filecolor=black,
               citecolor=black,
               urlcolor=blue
              }
     \newcommand{\web}[1]{\Blb{\url{#1}}}
\else
     \newcommand{\web}[1]{#1}
\fi

\newcommand{\ntab}[2]{ \multicolumn{1}{#1}{#2} }
\newcommand{\nntab}[2]{ \multicolumn{2}{#1}{#2} }
\newcommand{\nnntab}[2]{ \multicolumn{3}{#1}{#2} }
\newcommand{\nnnnnnnnntab}[2]{ \multicolumn{9}{#1}{#2} }

\definecolor{Dred}{rgb}{0.312,0.070,0.070}
\definecolor{Dblue}{rgb}{0.070,0.070,0.312}
\definecolor{Dgreen}{rgb}{0.070,0.312,0.070}
\definecolor{Db}{rgb}    {0.050,0.0,0.320}

\newcommand{\Blb}[1]{\textcolor{Dblue}{\bf #1}}

\newcommand{\beq}{ \begin{eqnarray} }
\newcommand{\eeq}[1]{\label{#1}\end{eqnarray}}
\newcommand{\eeqn}{ \nonumber \end{eqnarray} }
\newcommand{\Frac}[2]{\frac{\displaystyle\strut #1}{\displaystyle\strut #2} }

\renewcommand{\tau}{\uptau}

\newcommand{\lp}{ \left(  }
\newcommand{\rp}{ \right) }

\newcommand{\flo}[2]{\mbox{#1} \cdot 10^{#2}}

\newcommand{\der}[2] {\frac{ \partial #1 }{ \partial #2 } }

\newcommand{\hp}{\phantom{+}}
\newcommand{\hm}{\phantom{-}}
\newcommand{\Sum}{\displaystyle\sum}
\newcommand{\TEC}{\rm TEC}

\newcounter{note}
\let\oldmarginpar\marginpar
\renewcommand\marginpar[1]{\-\oldmarginpar[\raggedleft\footnotesize #1]%
{\raggedright\footnotesize #1}}
\newcommand{\Note}[1]{\Rdb{#1}{\addtocounter{note}{1}%
\marginpar{\small\underline{\Rdb{\footnotesize Corr \arabic{note}}}}}}
\newcommand{\note}[1]{\Rdb{#1}}
\renewcommand{\Note}[1]{#1}  
\renewcommand{\note}[1]{#1}  
\newcommand{\Number}[1]{\ifnum#1<10\relax0\number#1\else\number#1\fi}
\newcommand{\isodate}{
\count151=\time
\divide\count151 by 60
\count151=\count151
\multiply\count151 by 60
\count152=\time
\advance\count152 by -\count151
\divide\count151 by 60
\count152=\count151
\multiply\count151 by 60
\count153=\time
\advance\count153 by -\count151
\Number{\year}.\Number{\month}.\Number{\day}--\Number{\count152}:\Number{\count153}
}

\shorttitle{Single band VLBI absolute astrometry}
\shortauthors{Petrov}
\received{08 November, 2022}
\revised{08 February, 2022}
\accepted{}
\published{}
\submitjournal{Astronomical Journal}

\begin{document}
\title{Single band VLBI absolute astrometry}

\author[0000-0001-9737-9667]{Leonid Petrov}
        \affil{NASA Goddard Space Flight Center \\
        Code 61A, 8800 Greenbelt Rd, Greenbelt, 20771 MD, USA}


\correspondingauthor{Leonid Petrov} \email{Leonid.Petrov@nasa.gov}

\begin{abstract}
   The ionospheric path delay impacts single-band very long baseline 
interferometry (VLBI) group delays, which limits their applicability for 
absolute astrometry. I consider two important cases: when observations 
are made simultaneously at two bands, but delays at only one band are 
available for a subset of observations and when observations are made at 
one band design. I developed optimal procedures of data analysis 
for both cases using Global Navigation Satellite System (GNSS) ionosphere 
maps, provided a stochastic model that describes ionospheric errors, and 
evaluated their impact on source position estimates. I demonstrate that 
the stochastic model is accurate at a level of 15\%. I found that using 
GNSS ionospheric maps as is introduces serious biases in estimates of 
declinations and I developed the procedure that almost eliminates them. 
I found serendipitously that GNSS ionospheric maps have multiplicative 
errors and have to be scaled by 0.85 in order to mitigate the declination 
bias. A similar scale factor was found in comparison of the vertical total 
electron contents from satellite altimetry against GNSS ionospheric maps. 
I favor interpretation of this scaling factor as a manifestation of the 
inadequacy of the thin shell model of the ionosphere. I showed 
that we are able to model the  ionospheric path delay to the extent that 
no \Note{noticeable} systematic errors emerge and we are able to adequately 
assess the contribution of the ionosphere-driven random errors on source 
positions. This makes single-band absolute astrometry a viable option that 
can be used for source position determination.
\end{abstract}


\section{Introduction}

  Method of VLBI absolute astrometry involves observations of many active
galactic nuclei roughly uniformly distributed over the sky. Data analysis
of group delays derived from these observations is usually performed in the 
accumulative mode, i.e. all VLBI absolute astrometry and geodesy observations 
are processed in a single least square solution for estimation of source 
coordinates, Earth orientation parameters, stations positions, and nuisance 
parameters, such as atmospheric path delay in zenith direction and clock 
function. The errors in source position adjustments are \Note{mainly} due to 
the thermal noise and the inaccuracy of modeling path delay in the atmosphere. 
They vary in a wide range from 0.05 to 50 mas depending on source flux density, 
network geometry, and the number of observables, with 1~mas error being 
typical. \Note{The grid of sources which positions are determined from 
dedicated absolute astrometry observing campaigns forms the basis that makes 
possible differential astrometry that involves observations of pairs of 
targets and calibrators with known positions. This distinction 
is often blurred in literature for brevity. We should be aware of that 
differential astrometry, despite being capable to determine very 
precise position {\it differences}, in principle cannot provide positions 
more precise than positions of the calibrators determined with absolute 
astrometry and it inherits its random and systematic errors.
}

  Since the contribution of the ionosphere to group delay is reciprocal to the
square of the observing frequency, usually absolute astrometry experiments 
are performed at two widely separated bands, 2.3/8.4 or 
4.3/7.6~GHz \citep{r:wfcs}. Processing of dual-band data allows us to eliminate 
effectively the contribution of the ionosphere. However, there are situations 
when absolute astrometry observations are available only at one band. 

  We know that compared with dual-band observations, single band absolute 
astrometry observations are affected by the contribution of the path delay 
in the ionosphere. Two questions arise: 1)~which data analysis strategy does
provide source coordinate estimates with the lowest uncertainties and 2)~how 
to account for the contribution of errors in the ionosphere modeling to 
reported source coordinate uncertainties? The goal of this study is to provide 
answers to these questions. \Note{The dual-band group delays are considered the 
ground truth free from the impact of the ionosphere in the framework of this
study. I took several trial datasets of dual-band VLBI observations from 
twenty-four hour observing sessions and used them as a testbed. I dropped 
existing observations at the second band during data analysis and compared 
results of single-band data against the reference solution that used both 
bands. I should note that the analysis presented in this article is specific 
to source positions determined with absolute astrometry and is not applicable 
to position differences determined with differential astrometry that has its 
own error model.}

\section{VLBI dataset used for trials}

  The primary dataset used for analysis is 263 twenty four hour experiments 
at the Very Large Baseline Array (VLBA) network since April 1998 through 
May 2021. All these data are publicly available at US National Radio 
Astronomy Observatory (NRAO) archive\footnote{\web{https://data.nrao.edu}}. 
The dataset consists of 
observing sessions under regular VLBI geodesy program RDV \citep{r:rdv}, 
astrometric VCS-II program \citep{r:vcs-ii}, its follow-ups VCS-III and 
VCS-IV, and geodetic CONT17 campaign \citep{r:beh22}. The motivation of 
this choice is to have a long history of observations, a homogeneous 
network, and both short and long baselines. VLBI absolute astrometry 
programs at declinations above $-40^\circ$ are almost exclusively run 
with VLBA. Therefore, conclusions made from processing trial runs at 
the VLBA can be propagated directly to the past and future astrometry 
programs at that network.

\section{Modeling the ionospheric contribution to path delay}

\subsection{Dual-band observations}

  The impact of the ionosphere dispersiveness on fringe phase 
is reciprocal to frequency in the first approximation. Therefore,
fringe phase in channel $i$ in the presence of the ionosphere becomes
\beq
     \phi_i = 2\pi \tau_{p} \, f_0 + \tau_{g} \, (f_i - f_0) - \Frac{\alpha}{f_i},
\eeq{e:i1}
   where $\tau_p$ and $\tau_g$ are phase and group delays, $f_i$ is the frequency
of the $i$th spectral channel, $f_0$ is the reference frequency and
\beq
  \alpha = \Frac{e^2}{ 8\, \pi^2 \, c \, m_e \,  \epsilon_o }
              \lp \int N_v \, d s_1 - \int N_v \, d s_2 \rp,
\eeq{e:i2}
where $N_v$ --- electron density, $e$ --- charge of an electron, $m_e$ ---
mass of an electron, $\epsilon_o$ --- permittivity of free space, and $c$ --- 
velocity of light in vacuum. Integration is carried along the line of sight.
Having substituted values of constants \citep{r:klo96} and expressing the total 
electron contents along the line of sight $\int N_v \, d s$ in 
$\flo{1}{16}$ electrons/$m^2$ (so-called TEC units or TECU), we arrive to
$ \alpha = \flo{1.345}{9} \:\: \mbox{sec}$/TECU.

  Phase and group delay are computed using fringe phases $\phi_i$ with
weights $w_i$ using least squares. The result can be expressed analytically 
after some algebra:
\beq
   \tau_{\rm gi} = \tau_{\rm if} + \Frac{\alpha}{f_e^2} \, \TEC,
\eeq{e:i3}
  where $\tau_{\rm if}$ is the ionosphere-free group delay, $\TEC$ is
$\int N_v \, d s$ expressed in TEC units, and $f_e$ is the effective ionospheric
frequency
\beq
   f_e = \sqrt \Frac
                 { \Sum_i^n w_i \cdot
                          \Sum_i^n w_i ( f_i - f_0)^2  \: - \:
                      \lp \Sum_i^n w_i ( f_i - f_0 ) \rp^2 \,
                 }
                 {
                   \Sum_i^n w_i ( f_i - f_0 )
                   \Sum_i^n \frac{w_i}{f_i} \: - \:
                   \Sum_i^n w_i \cdot
                   \Sum_i^n w_i \frac{(f_i - f_0)}{f_i}
                 }
\eeq{e:i3a}
  that depends on weights of spectral channels $w_i$. Typically, the 
effective ionospheric frequency is within several percents of the 
central frequency of the observing band.

  The best way to mitigate the impact of the ionosphere on group delay is
to observe simultaneously at two or more widely separated frequency bands.
Then the following linear combination of two group delays at the upper
and lower bands, $\tau_{u}$ and $\tau_{l}$, respectively is 
ionosphere free:
\Note{
\beq
 \tau_{if} = \Frac{f_{u}^2}{f_{u}^2 - f_{l}^2} \, \tau_{u} -
             \Frac{f_{l}^2}{f_{u}^2 - f_{l}^2} \, \tau_{l}.
\eeq{e:i4}
}

  Here $f_{u}$ and $f_{l}$ are effective ionospheric frequencies at the upper
and lower bands respectively.

  The residual contribution of the ionosphere \Note{in dual-band combinations
is caused by systematic errors, namely} a)~higher order terms in the 
expansion of the dispersiveness on frequency \citep{r:iono2nd}; 
b)~the contribution of frequency-dependent source structure, and 
c)~the dispersiveness in the signal chain. These contributions affect group 
delay at a level of several picoseconds and they are considered insignificant 
with respect to other systematic errors.

\subsection{The use of TEC maps for computation of ionospheric path delay}

  TEC maps \citep{r:igs-gim} also known as Global Ionospheric Model (GIM)
derived from analysis of Global Navigation Satellite System (GNSS) data 
are used for reduction of single band observations. In particular, I used 
CODE TEC time series \citep{r:schaer99} available at 
\href{ftp://ftp.aiub.unibe.ch/CODE}{ftp://ftp.aiub.unibe.ch/CODE} 
since January~01 1995 with a spatial resolution of $5^\circ \times 2.5^\circ$. 
Time resolution varied. It was $24^h$ since 01 January 1995 through 
March~27, 1998; $2^h$ since March~28, 1998 through October~18, 2014; 
and $1^h$ after that date. The ionosphere is considered as a thin shell 
at the constant height $H_i$ of 450~km above the mean Earth's radius. 
The ionospheric contribution is expressed via TEC as
\beq
  \tau_i = \Frac{\alpha}{f^2_e} \: M(e) \: \mbox{TEC},
\eeq{e:i4a}
where $M(e)$ is the so-called thin shell ionospheric mapping function
\beq
     M(e) = \Frac{1}{\sqrt{1 - \lp \Frac{ \bar{R}_\oplus}
                                     {R_\oplus + H_i} \rp^2 \, \cos^2{ e_{\rm gc}} 
                          }
                     },
\eeq{e:i5}
  and $R_\oplus$ is the mean Earth's radius and $e_{\rm gc}$ is the geocentric
elevation angle with respect to the radius vector between the geocenter
and the station. Computation of TEC value at a given moment of time is 
reduced to computation of the position of the ionosphere piercing point at 
a given azimuth and elevation and interpolation of TEC at the piercing point
at that latitude, longitude, and time.

  TEC maps from GNSS is a coarse model of the ionosphere. Errors of $\tau_i$
computed according to expression~\ref{e:i4} are greater than the residual
ionosphere contribution of ionosphere-free linear combinations of group delays. 
Therefore, dual-band delay observables are preferred when they are available. 
However, there are two cases when they are not available: a)~dual-band observing 
sessions with some source detected only at one band; b)~single-band observing
sessions. In these two cases we resort to computation of the  ionospheric 
contribution to path delays using GNSS TEC maps and evaluation of uncertainties 
of these contributions.

\subsection{Ionospheric contribution in dual-band observing sessions when 
a source is detected at one band only}

  The simplest way to deal with a mixture of dual- and single- band data is to 
process experiments three times: 1)~using dual-band data of those observations that 
detected a source in both bands, 2)~using low band data, and 3)~using upper 
band data with applying ionospheric path delay computed from GNSS TEC maps. However, 
typically only a fraction 2 to 20\% of observations is detected at only one band; 
the rest of observations are detected at both bands. Therefore, we can use available 
dual-band observations at a given observing session to improve the TEC model.

  I represent ionospheric path delay at stations $j,k$ as 
\beq
   \begin{array}{lcl}
      \tau_i(t) & = & b_j(t) - b_k(t) \: + \: \\
                &   & \Frac{\alpha}{f^2_e} 
                      \biggl( \Bigl(\TEC_j(\phi_j,\lambda_j,t) + a_j(t) \Bigr) M(e_j) \: - \: \\
                &   & \phantom{aaaa} \Bigl(\TEC_k(\phi_k,\lambda_k,t) + a_k(t) \Bigr) M(e_k) \biggr),
   \end{array}
\eeq{e:i6}
   where $b_j(t) = \sum_i^n B^0_i(t) b_{ij}$ is a delay bias expanded over 
the B-spline basis of the 0th degree, $a_j(t) = \sum_i^n B^3_i(t) a_{ij}$ 
is the TEC bias expanded over the B-spline basis of the 3rd degree.
$\phi, \lambda$ are coordinates of the ionosphere piercing point that 
depend on positions of observing stations as well as azimuths and 
elevations of observed sources.

  The clock bias occurs due to path delay in VLBI hardware that 
is different at different bands. This bias is constant for most of 
the experiments, however occasionally breaks may happen at some stations.
Epochs of these breaks coincide with the epochs of breaks in the clock
function. Expansion over the B-spline basis of the 0th degree accounts for 
these breaks. (B-spline of the 0th degree is 1 within the range of knots
[i, i+1] and 0 otherwise.)

  I estimated parameters $a_j$ for all the stations and $b_j$ for all the 
stations except the one taken as a reference using all available dual-band 
observations of a given experiment using least squares with weights 
\beq
   w_i = \Frac{1}{\sqrt{y^2 + \frac{ f_{u}^4 \sigma^2(\tau_u) + 
                                     f_{l}^4 \sigma^2(\tau_l) }
                                   {(f_{u}^2 - f_{l})^2} }
                },
\eeq{e:i7}
  where $\sigma(\tau_u)$ and $\sigma(\tau_u)$ are group delay uncertainties
and $y$ is the error floor, 12~ps, introduced to avoid observations with 
very high signal to noise ratios to dominate the solution.

\begin{figure}
  \noindent\includegraphics[width=0.47\textwidth]{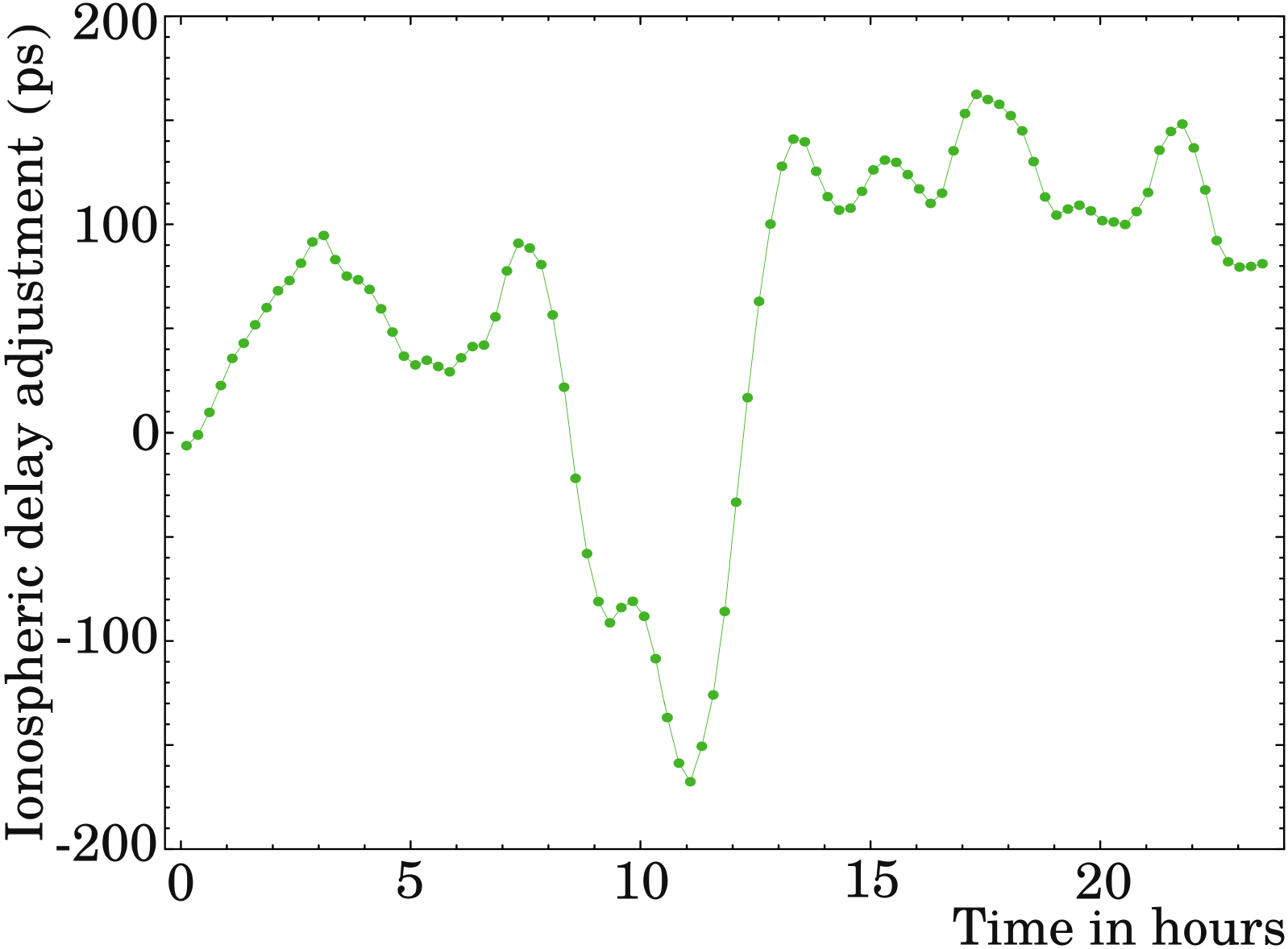}
  \caption{Adjustment to the ionosphere path delay bias at 8.4 GHz 
           with respect to the path delay derived from GNSS TEC 
           maps at {\sc mk-vlba} station from processing dual-band 
           observations on April 22, 2015. \note{The bias estimate
           corresponds to the averaged positions of the ionosphere 
           piercing points within a given time interval.} 
           1~TECU causes group delay 21~ps at 8.4~GHz.
          }
  \label{f:iono_bias}
\end{figure}

  The time span of the B-spline knot sequence for TEC bias in my solutions was
15~minutes. I applied constraints on the value of the B-spline coefficients and on
the first and the second derivatives with the reciprocal weights 
$5 \cdot 10^{-10}$ $s$, $4 \cdot 10^{-14}$, and $2 \cdot 10^{-18}$ $s^{-1}$ 
respectively. These constraints were introduced to ensure the continuity of 
biases and to prevent a singularity in rare cases when too few available 
observations at a given station could be used for bias estimation at a given 
spline segment. Figure~\ref{f:iono_bias} illustrates estimates of the 
ionospheric bias.

\begin{figure}
  \includegraphics[width=0.47\textwidth]{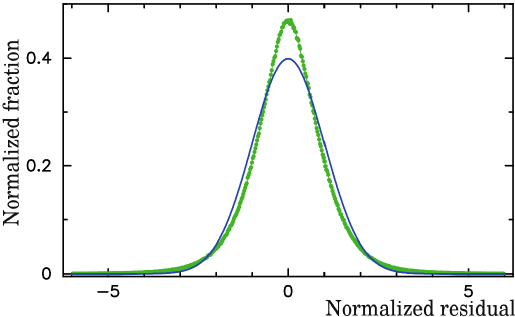}
  \caption{Empirical distribution of the normalized differences of 
           the ionosphere path delay computed from the GNSS TEC maps
           adjusted for clock and TEC biases (green dots). The 
           normal distribution with $\sigma=1$ is shown as a reference
           (solid blue line).
          }
  \label{f:iono_est_mod_dstr}
\end{figure}

  The resulting total electron contents model $\TEC_j(t) + a_j(t)$
is more precise than the a~priori $\TEC_j(t)$ taken from GNSS maps because
it uses additional information. Using estimates of $a_j$ and $b_j$ spline
coefficients, I compute $\tau_i(t)$ and its uncertainty according to the 
law of error propagation using the full variance-covariance matrix of
spline estimate coefficients. In order to evaluate the realism of 
these errors, I processed the trial dataset and computed $\tau_i(t)$ using
the estimates of clock and TEC biases and compared them with the
ionospheric contribution derived from dual-band observations.
I removed clock biases from VLBI dual-band ionospheric contributions
$\tau_{\rm vi}$, formed the differences $\tau_i - \tau_{\rm vi}$, and
then divided them by $\sigma(\tau_i)$ derived from the variance-covariance 
matrix of $a_j, b_j$. I generated the normalized histogram from the
dataset of 4,343,782 differences and computed the first two moments of 
the empirical distribution shown in Figure~\ref{f:iono_est_mod_dstr}. 
The fitting parameters of the first and second moments of the distribution 
are 0.003 and 0.889 respectively. Two factors cause a deviation of the 
second moment from 1.0 in the opposite way: a)~TEC variations not 
accounted by the parametric model; b) statistical dependence of the 
estimates of $a_j, b_j$, and VLBI path delay. After scaling the 
variance-covariance matrix by square of $0.889$, the distribution of the 
normalized residuals becomes close to Gaussian. The closeness of the 
empirical distribution to the normal distribution provides us 
a confidence that the extra noise introduced by the mismodeled ionospheric 
path delay after applying clock and TEC biases is properly accounted for. 

  The closeness of the distribution of normalized differences to Gaussian
with $\sigma=1$ is encouraging, but it does not guarantee that the residual
errors of the sum of TEC from GNSS and TEC bias adjustments cause no 
systematic error in estimates of source positions. To characterize the impact
of residual errors of the ionospheric contribution on source position,
I ran solution XIA that had the following differences with respect to the 
reference solution: 1)~it used X-band group delays; 2)~data reduction for the
ionosphere accounted for both a~priori TEC from GNSS maps and the ionosphere 
bias adjustment (expression in the parentheses \ref{e:i6}); 3)~errors 
of the ionospheric biases $\sigma  = a_{\rm ij} \, \alpha/f^2_e$ were added in 
quadrature to reciprocal weights of observables. \Note{It should be stressed
that parameterization, editing, and delay uncertainties were exactly the same 
as in the reference dual-band solution. In all trial solutions I varied only
input observables, ionosphere-specific reduction model and added noise to
the delay uncertainties. Since I analyzed only the differences in solutions,
the contribution of other model deficiencies, such as path delay in the neutral
atmosphere is canceled when I formed the differences. } 

  I ran also solution SIA that differed from XIA by using S-band group delays 
and the S-band effective ionospheric frequency $f_e$. For control, I ran 
solutions XIN and SIN that used the ionosphere-free combinations of group 
delays, the same weights as XIA and SIA solutions respectively, and the zero 
mean Gaussian noise with $\sigma = a_{\rm ij} \, \alpha/f^2_e$ was added to 
each observable. The differences in XIA and SIA solutions with respect to the 
reference solution provide us a measure of the impact of residual ionospheric 
errors on source position. The differences in XIN and SIN solutions 
characterize the impact of the residual ionospheric errors if they were 
Gaussian and totally uncorrelated.

\begin{figure*}
   \includegraphics[width=0.48\textwidth]{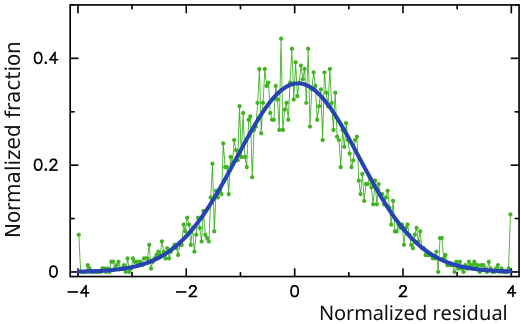}
   \hspace{0.03\textwidth}
   \includegraphics[width=0.47\textwidth]{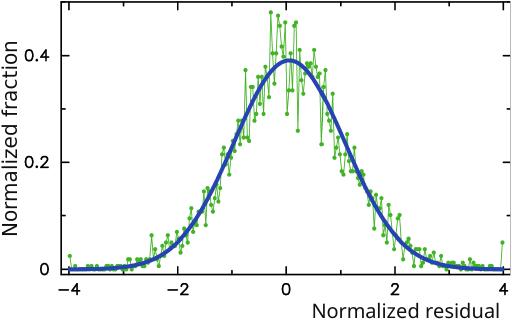}
   \caption{The distribution of normalized differences in declination
            from the trial VLBI solutions with respect to the 
            reference dual-band solution. {\it Left:} solution with
            GNSS TEC maps + adjustments of TEC biases using dual-band 
            VLBI group delays (XIA). {\it Right:} solution with added 
            Gaussian noise with $\sigma$ equal to errors in path delay
            that correspond to TEC bias adjustments uncertainties (XIN). 
            Blue thick lines show fitted Gaussian distributions with 
            the first and second moments listed in 
            Table~\ref{t:post_iono_resid}.
           }
   \label{f:post_iono_resid}
\end{figure*}

  Figure~\ref{f:post_iono_resid} shows the distribution of differences in 
declinations of source position estimates from XIA and XIN solutions. 
There is no noticeable deviation from the Gaussian shape. 
Table~\ref{t:post_iono_resid} lists first two moments of the distribution
of position differences in both right ascension and declination.
The second moment from SIN solution is close to 1.0, while the second moment
from XIN solution is 0.56. The errors of the ionospheric contribution at S-band 
dominate the error budget. These errors are 14 times less at X-band and are
only a fraction of overall group delay errors. The mean biases in right 
ascension and declination are negligible. The second moment of position 
estimates from XIA and SIA solutions is 15\% higher than moments from 
positions from XIN and SIN solutions respectively. This increase occurs due 
to non-randomness of residual ionospheric errors and can be viewed as a measure 
of unaccounted systematic errors in source positions due to the ionosphere. 
Analysis of these trial solutions demonstrates that we are able to predict 
the impact of ionospheric errors on source position with an accuracy 
of 15\% when TEC bias is estimated using dual-band observations.

\begin{table}[h]
   \begin{center}
      \begin{tabular}{|l|rr|rr|}
         \hline
          Parameter             & \nntab{l|}{TEC biases adjusted} & \nntab{l|}{Gaussian noise added} \\
                                & mean   & $\sigma$     & mean   & $\sigma$          \\
         \hline
         $\Delta \alpha$ X-band & -0.03  & 0.63         & -0.01  & 0.56              \\
         $\Delta \delta$ X-band &  0.02  & 0.64         &  0.01  & 0.56              \\
         $\Delta \alpha$ S-band & -0.05  & 1.14         & -0.01  & 1.03              \\
         $\Delta \delta$ S-band &  0.06  & 1.13         &  0.02  & 1.02              \\
         \hline
      \end{tabular}
   \end{center}
   \caption{The first and second moments of position differences of trial
            solutions XIA, SIA, XIN, and SIN with respect to the position
            estimates derived from the reference solution. The a~priori 
            TEC from GNSS and bias adjustment from dual-band solutions were
            applied in XIA and SIA solutions (the 2nd column). The zero-mean 
            Gaussian noise with $\sigma$ equal to the uncertainty in
            path delay from the TEC bias adjustment was added in XIN and 
            SIN solutions (the 3rd column).
           }
   \label{t:post_iono_resid}
\end{table}

\subsection{Ionospheric contribution in single-band observing sessions}

  TEC biases cannot be computed when an entire session is observed only 
at one band. Therefore, we have to resort to deriving a regression 
model to provide estimates of these errors. In the past, \citet{r:lcs2} 
derived a regression against the so-called global TEC: the integral of 
TEC over the entire Earth following ideas of \citet{r:gtec} and 
\citet{r:kra21} derived a regression against the root mean square (rms) of 
the total ionospheric path delay from GNSS TEC. In this study I use the 
second approach with slight modifications. Following general results of the 
turbulence theory \citep[see][]{r:tat71}, we can expect that fluctuations 
at scales $x$ are related to fluctuations at scales $y$ via a power law.

  I processed the same dataset of 263 twenty-four hour VLBI experiments 
that was used in the previous section and computed residual ionospheric
path delay for each observation as
\beq
   \tau_r = (\tau_{\rm gi} - \tau_{\rm vi} - (b_j - b_k)) \; \tilde{M},
\eeq{e:i8}
 where $\tau_{\rm gi}$ is the vertical ionospheric path delay from GNSS TEC 
maps, $\tau_{\rm vi}$ is the vertical ionospheric path delay from VLBI, 
$b_j - b_k$ are contributions of the clock bias, and $\tilde{M}$ is the 
averaged ionosphere mapping function between stations 1 and 2 of a given 
baseline: $\tilde{M} = (M(e_1) + M(e_2))/2.0$. The clock biases are routinely 
adjusted during analysis of VLBI observations and therefore, their contribution 
on VLBI results, such as source positions, is entirely eliminated. Subtracting 
them in expression \ref{e:i8}, I eliminate their impact on statistics as well. 
I used only twenty-four hour VLBI experiments for deriving statistics because 
the ionospheric path delay strongly depends on Solar time, especially at low 
latitudes, and statistics derived at shorter time intervals \Note{may not be}
representative.

\begin{table}
   \caption{Coefficients of the B-spline expansion of the 
            dependence of the rms of residual ionospheric path delay 
            derived from GNSS TEC maps on the rms of the total 
            ionospheric path delay from GNSS TEC maps at 8~GHz.
           }
   \begin{center}
      \begin{tabular}{crr}
        knot  index & knot argument  & B-spline value \\
                    & (ps)           & (ps)           \\
          \hline
             -2     &        &   6.3     \\
             -1     &        &  14.8     \\
          \hm 0     &        &  23.5     \\
          \hm 1     &    0.0 & 114.0     \\
          \hm 2     &   35.0 & 114.0     \\
          \hm 3     &  120.0 & 114.0     \\
          \hm 4     & 1300.0 &           \\
          \hline
      \end{tabular}
   \end{center}
  \label{t:iono_rms_rms}
\end{table}

\begin{figure}
  \includegraphics[width=0.47\textwidth]{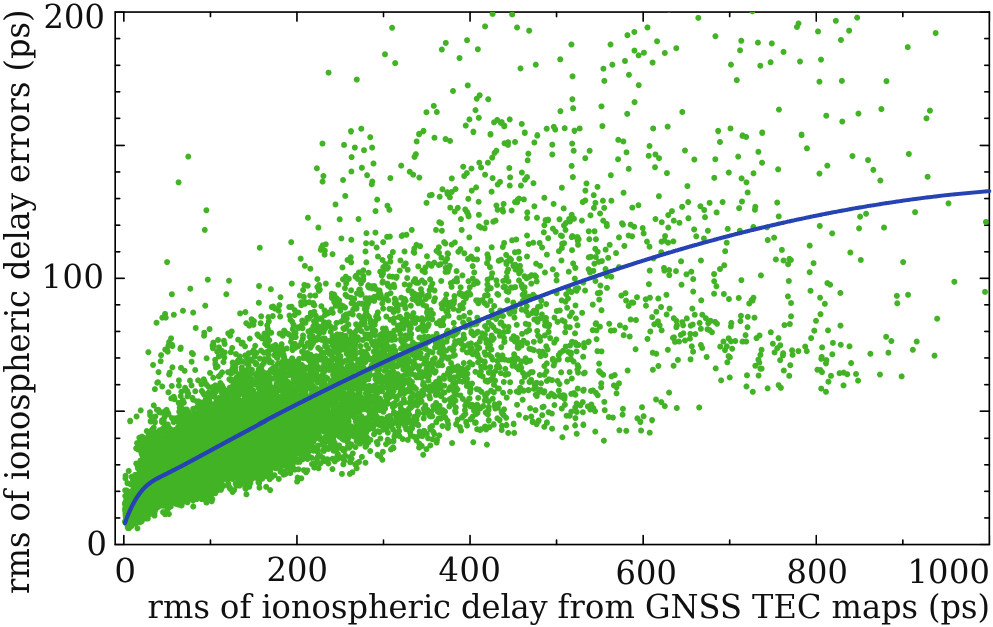}
  \caption{Dependence of the rms of residual ionospheric path delay derived
           from GNSS TEC maps $\sigma_{\rm gr}$ on the rms of the total 
           ionospheric path delay from these maps $\sigma_{\rm gt}$. 
           No adjustment to TEC has been applied. Path delay is computed 
           for the reference frequency 8~GHz. The blue smooth line shows 
           the regression model in a form of a B-spline that fits the data.
          }
  \label{f:iono_rms_rms}
\end{figure}

  Figure~\ref{f:iono_rms_rms} shows the dependence of the rms of residuals 
$\tau_r$ on the rms of total ionospheric path delay from GNSS TEC maps 
$\tau_{\rm gi}$. Each point on the plot corresponds to the rms for 
a given baseline and a given observing session. I confirm the early result of
\citet{r:kra21} but here I used a much larger dataset. The result reported
in \citet{r:kra21} was slightly affected by an error in computation of 
ionospheric group delays for a case when some data are flagged because of 
radio interference. This error has been fixed and the affected experiments 
have been reprocessed from scratches. This dependence can be coarsely described 
as a square root of the rms of the total ionospheric path delay. For a better 
approximation I sought a regression in the form of an expansion over 
B-splines of the 3rd degree. The spline coefficients computed using 
least squares are listed in Table~\ref{t:iono_rms_rms}.

  Using that regression, I developed the following algorithm for computation 
of errors of the ionospheric path delay from GNSS TEC maps. First, coordinates
of $K$ points uniformly distributed over the sphere are computed using a 
random number generator. Then for each baseline and each time epoch azimuth 
and elevation angles of that point are computed at both stations of the 
baseline, and if the elevations above the horizon are greater than $5^\circ$ 
at both stations, that point is selected for further computations. If not, 
the next point is drawn. Then total ionospheric path delay 
$\tau_i(A_1,e_1,A_2,e_2)$ is computed using GNSS TEC maps. It is worth 
mentioning here that unlike to troposphere path delay, 
$\tau_i(A,e) \neq \tau_i(A,\pi/2) \cdot M(e)$, since path delay depends 
on positions of the ionosphere piercing points. It is not sufficient to 
compute the ionospheric path delay in zenith direction and then map it via 
$M(e)$: latitude and longitude of the piercing point can be as far as 
1000~km from the station. Following this approach, we sample 
piercing points uniformly distributed within the mutual visibility zone. 
The process is repeated for 1440 time epochs that cover the time interval 
of VLBI experiment under consideration with a step of 1~minute. Then for 
each baseline $\sigma(\tau_{\rm gt})$ is computed over a time series of 
1440 $\tau_i$ values. Finally, the estimate of the rms of residual 
ionospheric path delay derived from GNSS TEC maps is computed from 
this regression via the rms of the total ionospheric path delay as
\beq
  \sigma_{\rm rr} = \sum_k^n B^3_k(\sigma(\tau_{\rm gt})) \, 
                    \sqrt{M^2(e_1) + M^2(e_2)}.
\eeq{e:i9}

  Baseline-dependent datasets are considered independent for this 
computation: the mutual visibility at all the stations of the network at 
a given moment of time is not enforced. For several baselines longer than 
96\% Earth diameter, this algorithm has a poor performance for 
selecting points above $5^\circ$. Therefore a minor modification 
is made for such an extreme case: the elevation angle is fixed to $5^\circ$, 
mutual visibility is not enforced, and azimuths are selected 
randomly within a range of $[0, 2\pi]$ independently for both 
stations. 

\begin{figure}
  \includegraphics[width=0.47\textwidth]{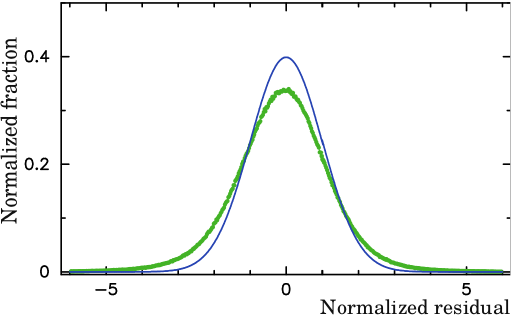}
  \caption{The distribution of the normalized differences of 
           the ionospheric path delay computed from the GNSS TEC maps
           against VLBI ionospheric path delay with clock biases 
           subtracted (green dots). The normal distribution 
           with $\sigma=1$  (solid blue line) is shown as 
           a reference.
          }
  \label{f:iono_regr_mod_dstr}
\end{figure}

  In order to evaluate the validity of this regression model of residual
ionospheric path delay errors, I computed $\tau_r$ from dual-band observations,
$\sigma_{\rm gt}$ following the algorithm described above for 263 
twenty-four experiments, and then computed the histograms of normalized 
residuals $\tau_r/\sigma_{\rm rr}$. The histogram is presented in 
Figure~\ref{f:iono_regr_mod_dstr}. The first  two moments of the distribution 
are $-0.083$ and 1.214 respectively. Since regression $\sigma_{\rm rr}$ 
was found using least squares, the number of observations with 
$\sigma(\tau_{\rm gr})$ less and greater than $\sigma_{\rm rr}$ for given 
$\sigma(\tau_{\rm gt})$ is approximately equal: the thick blue line cuts the 
cloud of green points in Figure~\ref{f:iono_regr_mod_dstr} almost by half. 
However, the variance of the contribution of those points with 
$\sigma(\tau_r) > \sigma_{\rm rr}$ overweights the contribution of those 
points with $\sigma(\tau_r) < \sigma_{\rm rr}$ because variance 
quadratically depends on $\tau_r$. This causes a positive bias. 
After multiplying $\sigma_{\rm rr}$ by 1.214, the distribution 
of normalized residuals becomes almost Gaussian.

  One can notice that $\sqrt{M^2(e_1) + M^2(e_2)}$ in expression
\ref{e:i9} is not the same as $\tilde{M} = (M(e_1) + M(e_2))/2$
used for computation of the regression. I found that using
$\tilde{M}$ instead of $\sqrt{M^2(e_1) + M^2(e_2)}$ decreases
the second moment from to 1.214 to 1.196, which is negligible,

\begin{figure}
   \includegraphics[width=0.47\textwidth]{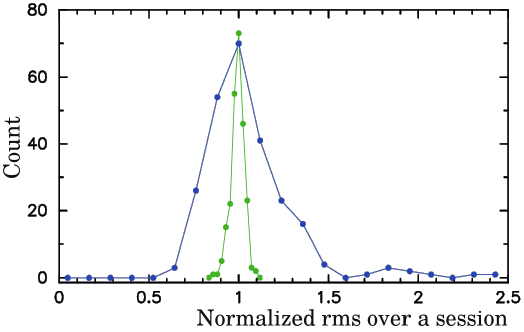}
   \caption{The distribution of second moment estimates of the 
            normalized differences of ionospheric path delays 
            derived from VLBI dual-band observations and GNSS TEC
            maps among individual observing sessions. The narrow green 
            curve shows the statistics of the normalized residuals 
            with TEC biases adjusted and the wide blue curve show 
            the statistics of normalized residuals without TEC adjustment.
           }
   \label{f:iono_mod_sess_nrml_dstr}
\end{figure}

  The distributions shown in Figures~\ref{f:iono_est_mod_dstr} and 
\ref{f:iono_regr_mod_dstr} are computed for the entire dataset of 
4.3 million path delays and they represent the general population 
over the interval of 23~years. Statistics for an individual
observing session may differ. In order to evaluate the scatter of
the statistics, I computed the time series of second moments of the 
distribution of normalized residuals of ionospheric path delays and 
their uncertainties with and without TEC biases adjusted for each 
observing session separately. I divided the normalized residuals by 
scaling factors of 0.889 and 1.196 respectively. I computed the 
distribution of second moment estimates and showed it in 
Figure~\ref{f:iono_mod_sess_nrml_dstr}. The scatter of the second 
moments is small when TEC biases are adjusted. That means this statistics
is robust. When TEC is not adjusted, the scatter is significantly
larger, but even in that case 90\% of the second moment estimates 
deviate from 1.0 by no more than 30\%. This provides us a measure of
uncertainties in computation of ionospheric path delay errors in
individual single-band observing sessions.

\subsection{The impact of the residual ionospheric errors in source position
in a case of single-band observing sessions}

  Error analysis presented in the previous section characterizes our ability
to predict the first and second moments of the distribution, but it does not
guarantee that residual errors due to the ionosphere cause no
systematic errors in source positions. I ran trial solutions XIT and SIT that
used X- and S-band group delays respectively, applied ionospheric path delays
derived from GNSS TEC maps, and inflated reciprocal weights of observables by
adding in quadrature errors in ionospheric path delays from TEC maps.

\begin{figure}
   \includegraphics[width=0.47\textwidth]{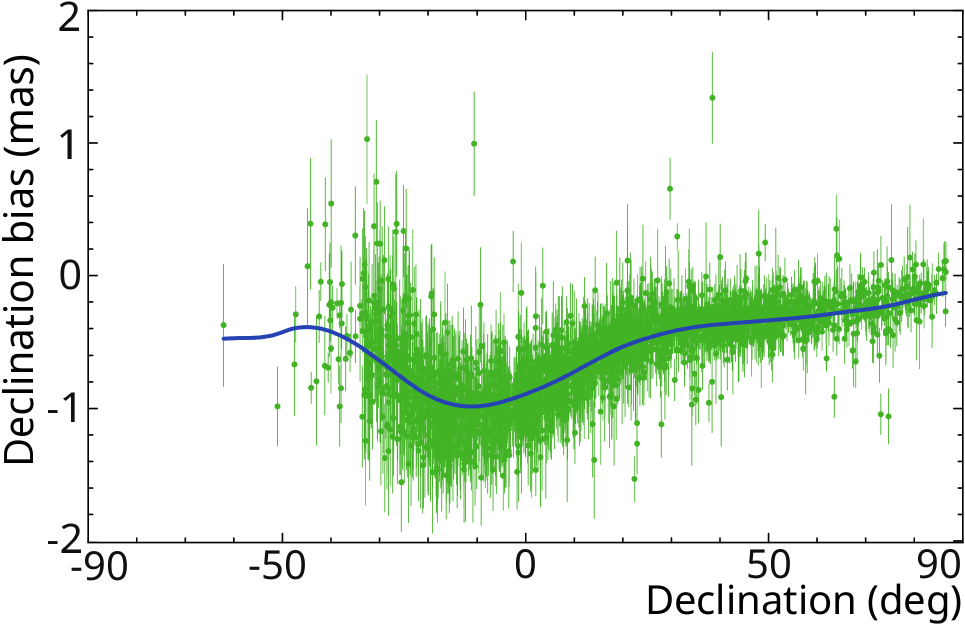}
   \caption{Differences in declinations from the X-band solution
            XIT using data at the VLBA network with the data reduction for 
            the ionosphere using ionospheric path delay from GNSS TEC maps 
            applied with respect to the dual-band reference solution. 
            The thick blue line shows the differences smoothed with the 
            Gaussian filter.
           }
   \label{f:x_bias_1000}
\end{figure}

\begin{figure}
   \includegraphics[width=0.35\textwidth]{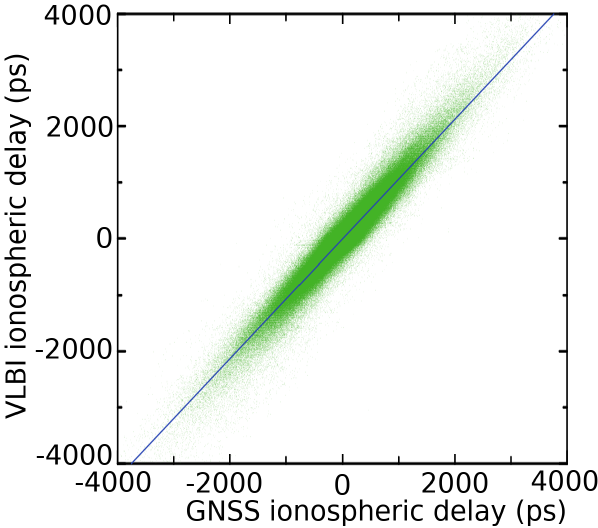}
   \caption{Dependence of VLBI ionospheric group delay at 8~GHz against
            the ionospheric group delay from GNSS. The blue straight line
            is the least square fit of this dependence.
           }
   \label{f:gv_reg}
\end{figure}

  Analysis of differences in source positions from trial XIT and SIT 
solutions with respect to the reference dual-band solution revealed no
peculiarities in right ascensions, but revealed significant systematic errors 
in declinations (see Figure~\ref{f:x_bias_1000}). The pattern of systematic 
errors in declination from the  S-band solution is similar but greater by 
a factor of $f^2_x/f^2_s \approx 14$. We cannot consider a solution with such 
errors as satisfactory. This was unexpected because prior work of 
\citet{r:sek03,r:hob06,r:det11,r:mot22} claimed a good agreement between TEC 
derived from VLBI and GNSS. And indeed, the plot of ionospheric contributions
in zenith direction from VLBI after removal clock biases against the ionospheric 
contributions from GNSS (Figure~\ref{f:gv_reg}) shows no peculiarities and fits 
the straight line $\tau_{\rm vi} = -3.9 \:{\rm ps} + 1.06 \cdot \tau_{\rm gi}$. 
Although the residuals of this dependence {\it look} random, they still cause 
systematic errors in source positions.

   I made a number of trial solutions. The following leads turned out productive:
1)~to modify mapping function and 2)~to scale ionospheric path delays from TEC.
\citet{r:schaer99} suggested the following modification of the ionospheric 
mapping function:
\beq
     M(e) = k \, \Frac{1}{\sqrt{1 - \lp \Frac{ \bar{R}_\oplus}
                                     {R_\oplus + H_i + \Delta H} \rp^2 \, \cos^2{ \alpha e_{\rm gc}} 
                          }
                     },
\eeq{e:i10}
   arguing that varying parameters $\alpha$ and $H_i$ one can
account for a more realistic electron density distribution with 
height than a thin shell model. Here $\Delta H$ is an increment in the 
ionosphere height and $\alpha$ is a fudge factor.

\begin{figure}
   \includegraphics[width=0.47\textwidth]{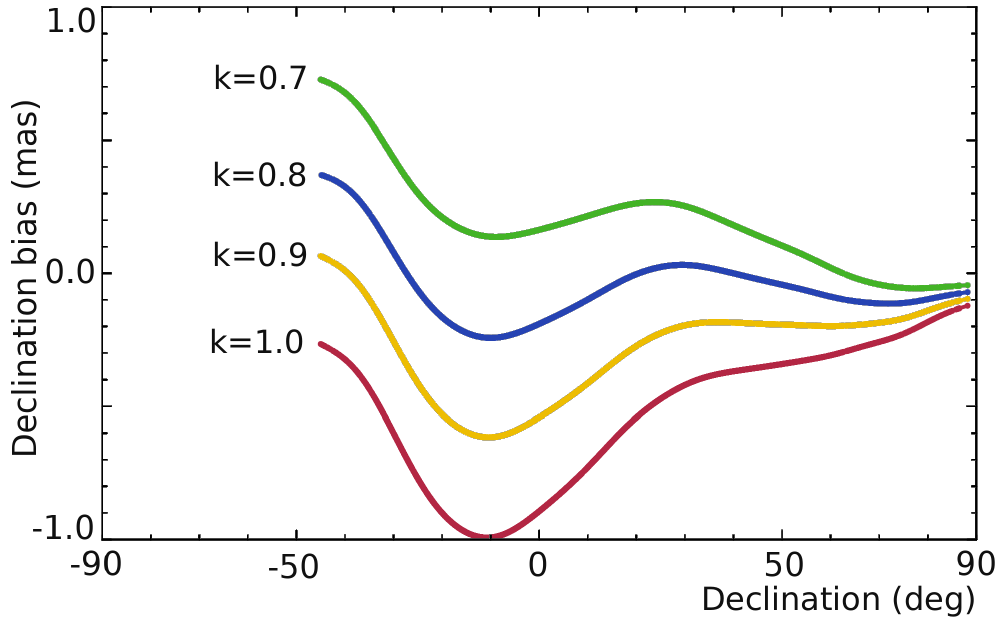}
   \caption{Smoothed declination biases from four solutions with respect 
            to the dual-band reference solution using X-band observables 
            and the ionospheric mapping function with parameters 
            $\Delta H=0$, $\alpha=1.0$ with different scaling factors 
            $k= 0.7,\, 0.8,\, 0.9,\, {\rm and} 1.0$.
           }
   \label{f:bias_4_map}
\end{figure}

\begin{figure}
   \includegraphics[width=0.47\textwidth]{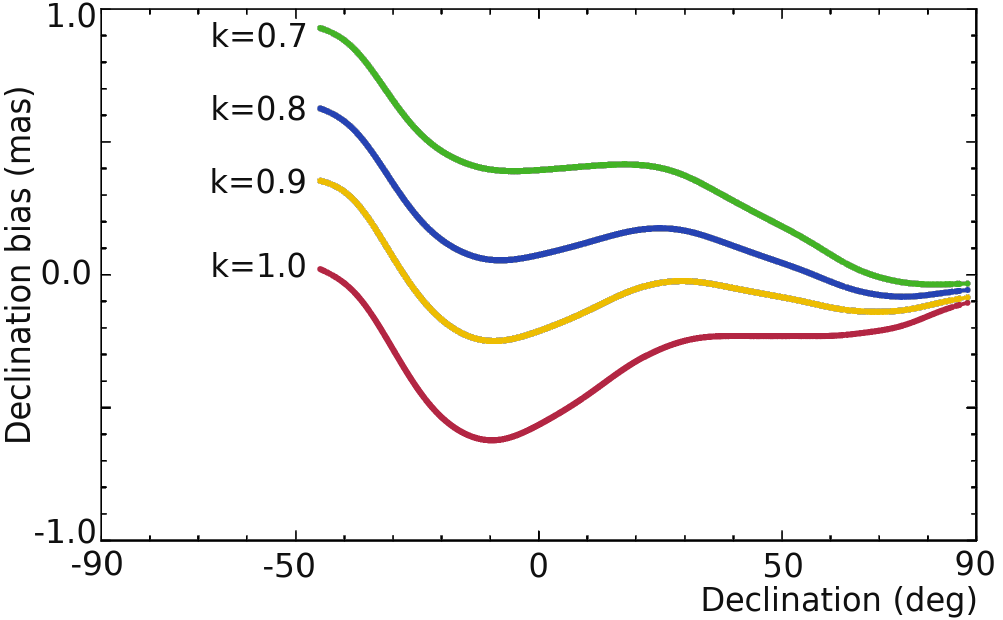}
   \caption{Smoothed declination biases from four solution with respect 
            to the dual-band reference solution using X-band observables 
            and the ionospheric mapping function with parameters 
            $\Delta H=56.7$~km and $\alpha=0.9782$ with different scaling factors 
            $k= 0.7,\, 0.8,\, 0.9,\, {\rm and} 1.0$. 
           }
   \label{f:bias_4_mod}
\end{figure}

\begin{figure}
   \includegraphics[width=0.47\textwidth]{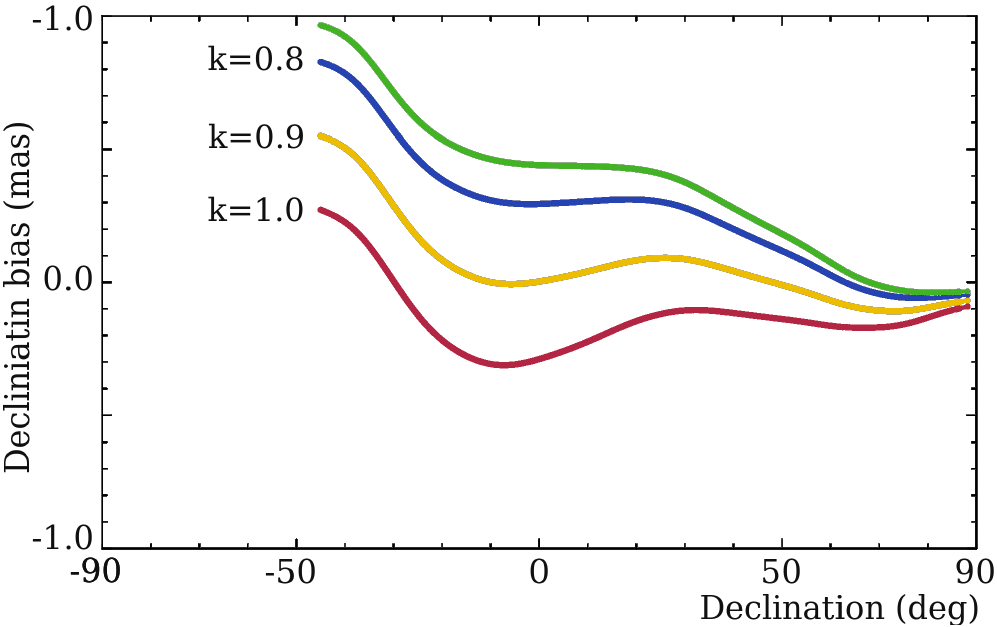}
   \caption{Smoothed declination biases from four solutions with respect 
            to the dual-band reference solution using X-band observables 
            and the ionospheric mapping function with parameters 
            $\Delta H=150$~km and $\alpha=0.9782$ with different scaling factors 
            $k= 0.7,\, 0.8,\, 0.9,\, {\rm and} 1.0$.
           }
   \label{f:bias_4_mhi}
\end{figure}

  I ran 12 trial solutions with mapping functions with parameters
1)~$\Delta H$=0.0,      \, $\alpha=1.0$; 
2)~$\Delta H$=56.7~km,  \, $\alpha=0.9782$; and
3)~$\Delta H$=150.0~km, \, $\alpha=0.9782$. 
Variant~2 corresponds to the so-called JPL modified single layer ionospheric 
mapping function. It was used in a number of papers 
\citep{r:li19,r:xiang19,r:shao21,r:wie21}, however no details how these 
values of $k,\, \Delta H,$ and $\alpha$ were derived have been provided. 
I varied the mapping function scaling factor $k$, setting it to 
0.7, 0.8, 0.9, and 1.0 and computed declination biases with respect 
to the reference solution. The results are presented in 
Figures~\ref{f:bias_4_map}--\ref{f:bias_4_mhi}.
  
  The declination bias is reduced when the parameter ionosphere height in 
the mapping function is increased. I interpret this result as a deficiency of 
GNSS TEC maps and I associate the origin of this deficiency with an 
over-simplification of the mapping function that was used for derivation of 
the TEC maps from processing of GNSS observations.

  Close analysis of these figures reveals that the declination bias has 
three components: 1)~a constant; 2)~a linear increase in the declination 
bias with a decrease of declination; 3)~a feature 
$\delta (x \, \sin \delta + y \, \cos \delta)$ with the minimum at 
approximately $-12^\circ$, where $\delta$ is declination. All three 
components depend on parameters of the mapping function
$\Delta H,\, \alpha$ and on scaling factor $k$. Selecting \Note{optimal}
$\Delta H,\, \alpha,\, k$, we can reduce declination biases. I selected
$\Delta H = 56.7~{\rm km},\, \alpha=0.9782,$ and $k=0.85$. This choice 
makes the weighted mean declination bias over all sources 0.013~mas. 
There is an element of subjectivity in this specific choice, since
there exist another combinations of $\Delta H, \alpha, k$ that
provide the zero mean bias. \citet{r:wie21} showed that the mean rms
of the difference between GNSS TEC maps and GNSS slant TEC was reduced
from 2 to 6\% when this modified mapping function was used.
That choice of $\Delta H$ and $\alpha$ led me to a selection of 
the specific scaling factor $k$.

\begin{figure*}
   \includegraphics[width=0.47\textwidth]{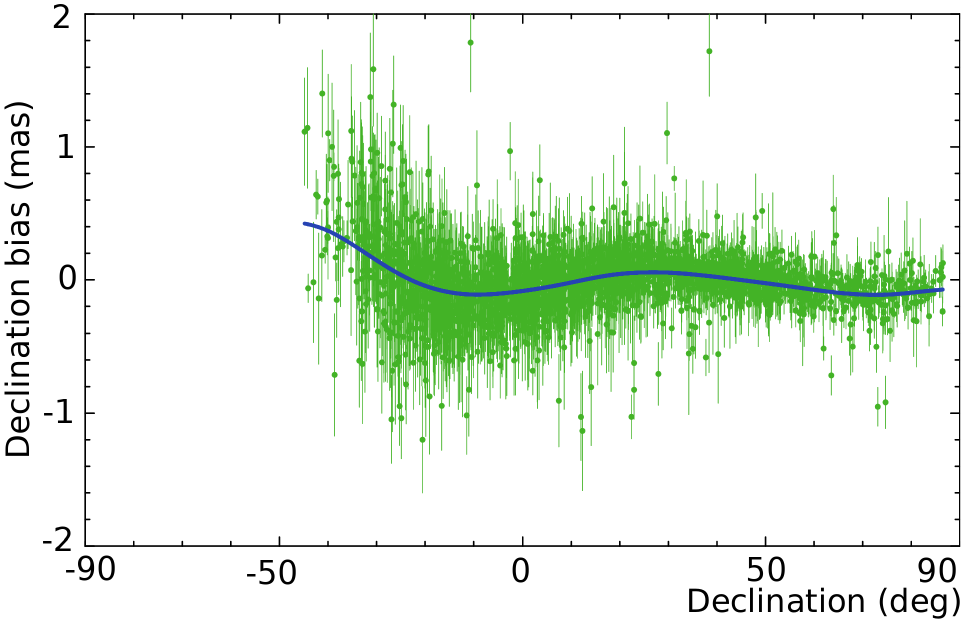}
   \hspace{0.03\textwidth}
   \includegraphics[width=0.47\textwidth]{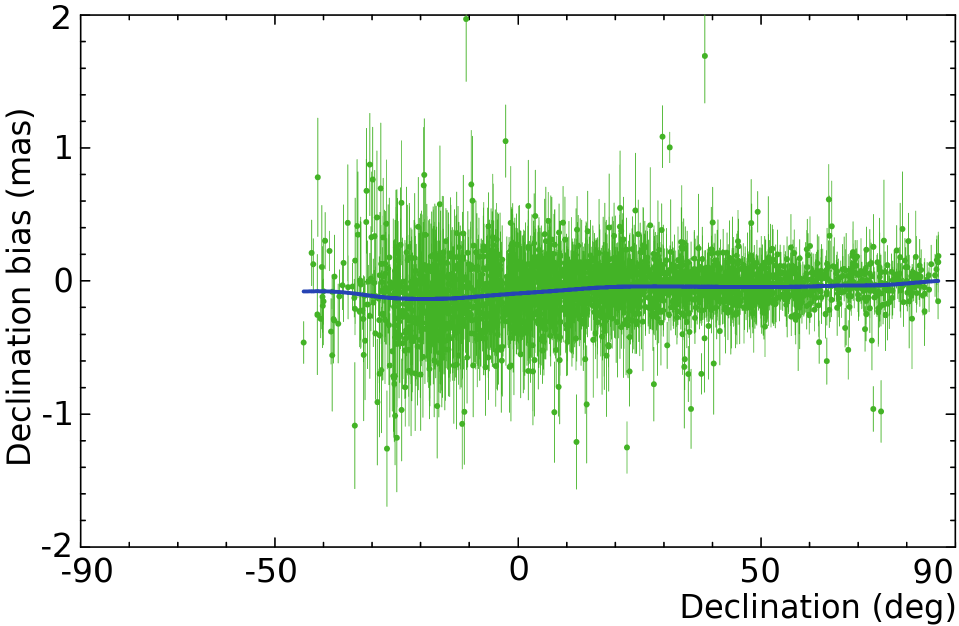}
   \caption{The declination bias with the ionospheric path delay
            using mapping function with parameters 
            $\Delta H=56.7 {\rm km}, \alpha =0.9782, k=0.85$
            with ({\it Right}) and without ({\it Left}) applying
            the empirical de-bias correction.
           }
   \label{f:debias_corr}
\end{figure*}

  Although scaling and modification of the ionospheric mapping 
function results in a substantial reduction of the declination bias, 
the remaining bias is still worrisome. To mitigate the bias even 
further, I introduce the ad~hoc correction for the ionospheric bias 
in the data reduction model. I smoothed the biases $D(\delta)$ with 
the Gaussian filter with $\sigma=8^\circ$ --- these smoothed biases 
are shown in Figures~\ref{f:bias_4_map}--\ref{f:bias_4_mhi} --- 
and expanded them over the basis of B-splines of the 3rd degree 
with 13  equi-distant knots in the range of $[-45^\circ, 90^\circ]$. 
\Note{The expansion coefficients are presented in the appendix.} 
I added correction
\beq
   \der{\tau}{\delta} \, D(\delta) / f^2_e
\eeq{e:i11}
  to observables. Here $f_e$ is the effective ionospheric path delay 
of a given observation. Figure~\ref{f:debias_corr} shows the effect 
of applying the de-bias correction. The bias has gone. 

   I ran two solutions using X-band only data (XIB) and S-band only
data (SIB), applied both the a~priori ionospheric path delay derived 
from GNSS TEC maps using the modified mapping function with 
parameters $\Delta H=56.7 {\rm km},\, \alpha =0.9782,\, k=0.85$
and the declination bias correction. The reciprocal weights 
of observables were adjusted by adding in quadrature the errors of 
residual ionospheric errors $\sigma_{\rm rr}$ modeled according to 
regression expression \ref{e:i9}. The first and the second moments 
of the normalized differences in sources positions are presented in 
Table~\ref{t:post_iono_sba}. The second moments are less than 1.0 
in the X-band solution. This indicates that the ionospheric 
contribution is not the dominating error source 
\Note{in these solutions.}

\begin{table}[h]
   \begin{center}
      \begin{tabular}{|l|rr|rr|}
         \hline
          Parameter      & \nntab{l|}{ X-band solution} & \nntab{l|}{S-band solution} \\
                         & mean   & $\sigma$  & mean   & $\sigma$ \\
         \hline
         $\Delta \alpha$ & -0.07  & 0.45      & -0.11 & 0.82  \\
         $\Delta \delta$ & -0.02  & 0.50      & -0.06 & 1.00  \\
         \hline
      \end{tabular}
   \end{center}
   \caption{The first and second moments of position differences of trial
            solutions XIB and SIB with respect to the source position
            estimates derived from the reference solution. The a~priori 
            ionospheric contribution was computed from GNSS TEC maps using 
            the modified mapping function with parameters
            $\Delta H=56.7 {\rm km},\, \alpha =0.9782,\, k=0.85$, the 
            declination bias correction was applied, but no reduction for
            TEC bias adjustment has been applied.
           }
   \label{t:post_iono_sba}
\end{table}

  Unfortunately, it does not look possible to {\it fix} the deficiency of 
the GNSS TEC maps without re-processing GNSS observations, which is well 
beyond the scope of this article, but nevertheless, it is still feasible
to {\it mitigate} the impact of the imperfection of GNSS TEC maps on 
source position estimates.

\section{Discussion}

\subsection{Inaccuracy of GNSS TEC maps}

  A number of authors compared the ionospheric contribution from GNSS,
VLBI, DORIS, radio occultation observation, and dual-band satellite altimeters
\citep{r:sek03,r:hob06,r:det11,r:para17,r:cokrlic18,r:li18,r:li19, 
r:xiang19,r:wie21,r:shao21,r:mot22}. A message these publications 
convey is there is a reasonable agreement between GNSS TEC maps 
and other techniques and there are no majors problems. Therefore, large
systematic errors driven by mismodeling of the ionospheric contribution
derived from GNSS maps came as a surprise. Do results presented in 
this study contradict to prior publications?

  Comparison in the past was often performed using very short continuous
dataset of VLBI observations, from 5 to 15 days 
\citep{r:hob05,r:det11,r:ete21,r:mot22}. The level of the agreement was 
characterized in terms of additive errors of GNSS TEC model. Unfortunately, 
when data from a short time interval are analyzed, a distinction between 
additive and multiplicative errors becomes blurry. Characterizing the 
differences in terms of biases and rms of their scatter did not turn out
productive in revealing systematic errors. Moreover, an unconscious 
bias of focusing a study on assessment of an agreement rather than 
investigation of disagreements, that were noticed and just briefly 
reported, diverted attention from investigating the differences in depth.

  However, reading between lines of published papers, we can find pieces
of evidences supporting findings in this work. The distribution of differences
in VLBI vertical TEC with respect TEC derived from GNSS TEC maps presented in
Figure~6 in \citet{r:hob06} shows a very significant skew. The distribution 
has a negative mean and a much greater left tail than the right tail. The
VLBI vertical TEC from that study appeared less than the TEC from GNSS TEC 
maps, in agreement with what I have found. The authors did not investigate 
that stark deviation of the distribution of over one million differences 
from the Gaussian shape, only noting that the bias is less than 3~TECU. 
Considering the errors are normally distributed and {\it additive}, and 
considering that the derivation of TEC from VLBI does not introduce serious 
errors, one can expect to arrive to the Gaussian distribution of residuals. 
For instance, the distribution of differences in 
Figure~\ref{f:iono_regr_mod_dstr} indeed does not show any measurable 
deviation from the Gaussian shape. However, if errors are {\it multiplicative}, 
the residuals will be non-Gaussian and their distribution will be skewed.

  Satellite altimetry using Jason satellites 
\citep[][and references therein]{r:jason} provides an independent way for
assessment of a level of the disagreement between direct vertical TEC 
measurements and GNSS TEC maps. \citet{r:li18} showed that comparisons of 
the differences revealed significant systematic biases that depend on 
geomagnetic latitude. However, an attempt to characterize these additive 
biases in terms of the rms was not very productive. \citet{r:liu18} presented 
spatial distribution of the differences (Figures 6--7). That distribution 
strikingly reminds the average distribution of TEC itself, suggesting the 
differences are multiplicative. A more recent paper of \citet{r:det22} 
revealed a highly significant scaling factor between GNSS TEC maps and 
four altimeter missions. The scaling factors varied from 0.809 to 0.919 
which is very close to what has been found in my analysis of astrometric 
observations of extragalactic radio sources. 

  \citet{r:li19} performed a comparison of TEC maps with Jason satellite
altimetry and with ionospheric radio occultations \citep[see the overview of
this technique in][]{r:ro}. They characterized the differences as 
a superposition of an additive bias and a multiplicative scale factor. 
The scale factors defined as 
TEC${}_{\rm Jason/GIM}$ and TEC${}_{\rm COSMIC}$/TEC${}_{\rm GIM}$ 
vary  with time and latitude and stay in a range of 0.7--0.9 at night and 
0.9--1.0 during daytime. Since Jason orbits have altitude 1,350~km and
GNSS orbits have altitude about 20,000~km, Jason altimetry misses the 
contribution from the upper layers of the ionosphere and the plasmosphere. 
\Note{\citet{r:yiz08} analyzed TEC between Jason and GNSS satellites
and found that the share of plasmosphere to TEC is on average 15\%
and may reach 60\% when TEC is low. However, such large share of the 
electron density at altitudes above 1,350~km is inconsistent with 
the thin shell model at the altitude of 450 or 506.7~km that assumes
no electron density above that height at all.}


\begin{figure*}
   \includegraphics[width=0.47\textwidth]{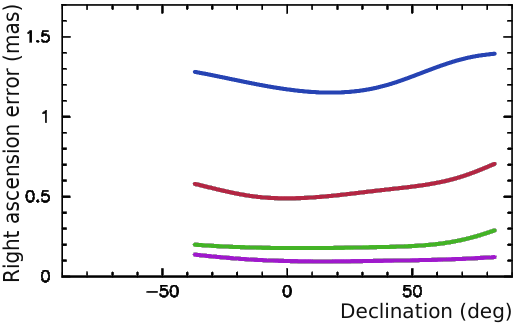}
   \hspace{0.05\textwidth}
   \includegraphics[width=0.47\textwidth]{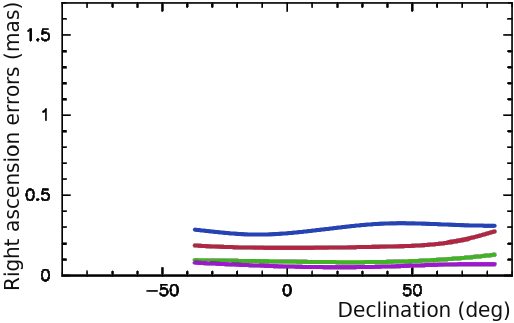}
   \includegraphics[width=0.47\textwidth]{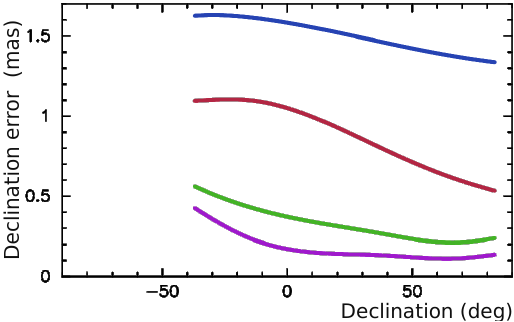}
   \hspace{0.05\textwidth}
   \includegraphics[width=0.47\textwidth]{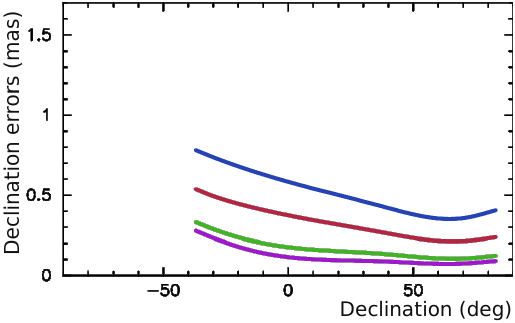}
  \caption{The rms of the differences in right ascensions ({\it upper row})
           and declinations ({\it lower row}) when single band group 
           delays are used with respect to the positions from the dual-band
           reference solution. A~priori ionospheric path delay using GNSS TEC 
           with the modified mapping function ($\Delta H=56.7~{\rm km}, 
           \alpha=0.9782, k=0.85$) was used adjusting TEC biases from
           dual-band observations ({\it right}) and without adjusting biases
           but with applying the de-bias correction from expression~\ref{e:i11}
           ({\it left}). The upper blue band shows the differences for 
           S-band, next red line shows the differences for C-band, next 
           green line shows the differences for X-band, and the bottom 
           purple line shows the differences for K-band.
           }
   \label{f:err_src_pos}
\end{figure*}

  It is essential to note that analysis of Jason and COSMIC data provides
estimates of vertical TEC, while analysis if GNSS observations provides slant 
TEC that is converted to vertical TEC using the mapping function. The dependence
of the declination biases on the mapping function found in this study
suggests that a simple thin model may not be adequate. \citet{r:schaer99} 
discussed the dependence of the effective ionosphere height on solar zenith 
angle. \citet{r:xiang19} studied it in more details. The height 
of the peak electron density has annual and diurnal variations. The latter
variations have an amplitude of about 100~km, being lower at daytime.
They showed that the instantaneous mapping function that accounts for 
the height of the electron content maximum achieves 8\% reduction of 
mapping errors. It should be mentioned that if the effective height of 
the ionosphere is changed, the latitude and longitude of the ionosphere 
piercing point for an observation with a given elevation and azimuth will
change as well, which will provide an additional change in path delay.

  These works strengthen argumentation in favor of that the thin shell 
model used for derivation of TEC maps since mid 1990s is oversimplified. 
The effective height of the ionosphere changes with time and latitude
and therefore, a realistic mapping function should also vary with time and 
latitude. Omission of this complexity results in a deficiency of GNSS TEC 
maps that manifests in multiplicative (scale) and additive (bias) errors 
that varies with time and latitude. The non-linear dependence of the 
declination bias with declination in a form of 
$\delta (x \, \sin \delta + y \, \cos \delta)$ may be caused by the 
dependence of GNSS TEC scaling factor with latitude reported by 
\citet{r:li19}. Radio waves from Southern Hemisphere sources observed at 
the array located in the Northern Hemisphere propagate through regions 
in the ionosphere with systematically lower latitudes than from Northern 
Hemisphere sources. Since a fixed mapping function is used for both 
computation of TEC maps and computation of slant ionospheric path delay 
from these maps, no modification of that fixed mapping function for data 
reduction of astronomical data is able to account for this kind of 
complexity, but as it was shown earlier, it is still possible to mitigate 
it. The remaining bias can be eliminated by applying the empirical de-bias 
correction.

\subsection{Omitted refinements of the ionosphere contribution}

  \citet{r:ete21} processed 60 days of VLBI data and estimated not only TEC
for each site using B-splines of the 1st degree, but also TEC derivatives
over longitude and latitude that were considered constant over 24~hour periods.
They claim that estimating TEC partial derivatives decreases discrepancies
of TEC from VLBI to GNSS TEC maps by 36\%. Inspired by this results, I 
introduced estimation of latitude and longitude TEC gradients in a form of
B-spline in addition to TEC estimation using dual-band data. I considered 
the weighted rms (wrms) of the differences between the parameterized model of 
TEC adjustment to vertical TEC from dual-band data as a metric of an 
improvement. I did not find a reduction of the rms greater then several 
percents and abandoned this approach. I should note this result does not 
disprove findings of \citet{r:ete21} since they used a different metric.

  Expression for the ionospheric contribution \ref{e:i1} is an approximation.
\citet{r:iono2nd} considered the impact of the higher order of expansion on 
group delay, namely proportional to $f^{-3}$. They found that the maximum
contribution of the 3rd degree term varied from 3 to 9~ps at 8~GHz depending 
on the baseline. This contribution is approximately one order of magnitude 
less than the contribution of residual ionospheric errors after taking
into account GNSS TEC maps. It should be noted that the 3rd degree term affects
both single-band and ionosphere-free linear combinations of group delays at 
two bands and therefore, cannot be retrieved in analysis of the differences 
with respect to the dual-band reference solution. 
 
\subsection{The impact of remaining errors in modeling of the 
ionosphere on source positions}

\begin{figure*}
   \includegraphics[width=0.49\textwidth]{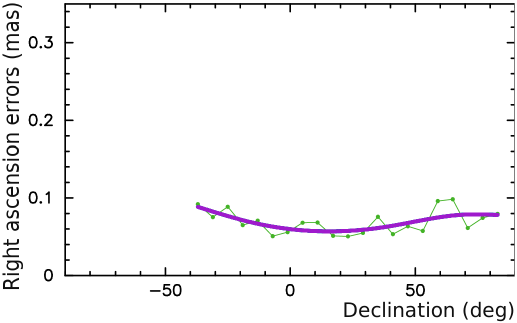}
   \hspace{0.01\textwidth}
   \includegraphics[width=0.49\textwidth]{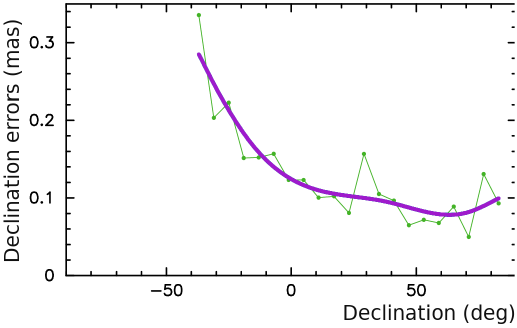}
   \caption{An increase in source position errors in right ascension 
            ({\it Left}) and declination ({\it Right}) due to the residual 
            ionospheric contribution at K-band (22~GHz) after applying 
            data reduction based on GNSS TEC maps with the modified mapping 
            function ($\Delta H=56.7~{\rm km}, \alpha=0.9782, k=0.85$) 
           }
   \label{f:k_err}
\end{figure*}

  As it was shown before, our ability to model the ionospheric contribution
using single-band observations is limited. After applying the declination
de-bias data reduction, the declination bias is virtually eliminated. The 
residual ionospheric contribution causes additional random errors. In order 
to investigate its impact, I ran four trial solutions. I used 
ionosphere-free linear combinations of dual-band data and added to them 
the ionospheric contribution derived from VLBI data scaled to the specific 
frequency of the trial solution. Then I applied the data reduction for the 
ionosphere to this dataset of modified observables using GNSS TEC maps as 
if I processed single-band observations and estimated source positions. 
The differences in source positions from these solutions with respect to 
the reference solutions are interpreted as an impact of the residual 
ionospheric errors at different frequencies. The declination dependence 
of the differences in right ascensions and declinations for four solutions 
for frequencies 2.3~GHz (S-band), 4.3~GHz (C-band), 8.4~GHz (X-band), 
and 23.7~GHz (K-band) is shown in Figure~\ref{f:err_src_pos}.

  These plots help to quantify additional errors in source positions that would 
arise if the dataset of 263 twenty-four hour experiments used for this study 
would have been observed at one band only. We see that estimation of TEC biases 
reduces the ionosphere-driven errors by a factor of 2--4. Still, even after 
estimation of TEC biases, the declination dependence of the ionosphere-driven 
errors remains.

  It is known that unaccounted source structure affects source position 
estimates. In general, the contribution of source structure at higher 
frequencies is less because jets become optically thin. Therefore, there was an 
expectation that observing at high frequencies, such as 22--24~GHz (K-band) one 
may obtain more precise source positions. So far, observational evidence do not 
support that prediction \citep{r:kq_astro,r:icrf3}. A detailed comparison of 
K-band absolute astrometry versus dual-band astrometry of \citet{r:kar19} did 
not reveal an improvement. Therefore, it is instructive to see what is the 
contribution of the ionospheric errors at K-band after applying data reduction 
from GNSS TEC maps. Figure~\ref{f:k_err} shows additional errors in right 
ascensions and declinations due to residual contribution of the ionosphere. 
We see that additional errors in right ascension are about 0.1~mas. Declination 
errors of Northern Hemisphere sources are at the same level. However, these 
errors grow with a decrease of declination for sources in the Southern Hemisphere 
approximately linearly and reach 0.3~mas at declination $-40^\circ$.
Therefore, the unmodeled ionospheric contribution sets the error floor in
position accuracy. This error floor is not accounted in the ICRF3 catalogue,
and the unaccounted ionospheric contribution is 300\% greater than the noise 
floor at K-band adopted by \citet[]{r:icrf3} according to their Table~6. 
The potential of K-band astrometry cannot be utilized unless the accuracy 
of ionosphere modeling will be substantially improved.

\subsection{The rms of residual atmospheric errors}

\begin{figure}
   \par\bigskip\par
   \includegraphics[width=0.47\textwidth]{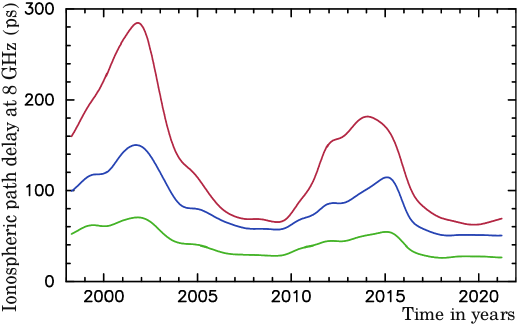}
   \caption{The rms of the mean residual ionospheric contribution
            at VLBA baselines at 8~GHz for three cases: 
            1)~no a~priori ionospheric contribution is applied (upper red curve); 
            2)~the a~priori ionospheric path delay computed using GNSS TEC maps 
               is applied (middle blue curve); and 
            3)~the a~priori ionospheric path delay computed using GNSS TEC 
               maps are applied and TEC biases are adjusted (lower green curve).
            1~TECU causes group delay 21~ps at 8~GHz.
           }
   \label{f:iono_mod_stat}
\end{figure}

  I investigated how the rms of the residual ionospheric contribution
for 45 VLBA baselines from the dateset of 263 twenty-four observing 
sessions varied with time. I computed three statistics 
for each observing session: 1)~$\tau_v - b_i$, 
2)~$\tau_v - \tau_g - b_i$; and 3)~$\tau_v - \tau_g - a_i - b_i$, 
and re-scaled them to 8~GHz. Here $\tau_v$ and $\tau_g$ are 
ionospheric path delays from VLBI and TEC maps respectively, $a_i$ 
is the adjusted TEC bias, and $b_i$ is clock bias. The statistics 
are shown in Figure~\ref{f:iono_mod_stat}. In order to improve 
readability, the time series were smoothed using Gaussian 
kernel with parameter $\sigma = 1$ year. The first statistics 
characterizes the impact of the total ionosphere on group delay. 
The second statistics characterizes the rms of the impact of the residual 
ionospheric errors on group delay after applying the a~priori GNSS TEC 
model. The final statistics characterizes the impact of residual
errors after applying the a~priori GNSS TEC and adjusting TEC biases
using dual-band VLBI data.

\subsection{Processing geodetic data from southern and R1/R4 VLBI experiments}

\Note{
   It would be  instructive to explore to which extent the numerical values 
   found in processing data at VLBA network are representative to experiments 
   conducted at other VLBI networks and to which extent they are specific
   to VLBA. I ran two solutions XIS and XIR. Solution XIS used 63,984 X-band 
   group delays from 36 VLBI experiments in 2016--2019 for absolute astrometry 
   programs at core Southern Hemisphere stations {\sc hart15m, hartrao,
   hobart12, hobart26, kath12m,  tidbin64, wark12m, yarra12m} with 
   participation of other International VLBI Service for geodesy (IVS) stations.  
   Solution XIR used 6,600,959 X-band group delays from 2154 geodetic VLBI 
   experiments from IVS programs R1 and R4 \citep{r:r1r4_proc} for 2002--2022. 
   Data from R1, R4, and Southern Hemisphere VLBI experiments are available at 
   the NASA Crustal Dynamic Data Information 
   System\footnote{\web{https://cddis.nasa.gov/archive/vlbi}}.
   In total, 36 globally distributed stations participated in experiments under 
   R1 and R4 programs. These experiments were designed for Earth orientation 
   parameter determination. I computed also reference solutions RIS and RIR 
   that used ionosphere-free combinations of dual-band group delays. Solution 
   parameterization and editing were identical for pairs of XIS/RIS and 
   RIS/RIR solutions.
}

\Note{
   I compared source positions from XIS/RIS and XIR/RIR solutions. In both
   cases declination systematic errors are evident. I found that scaling 
   that makes the mean declination bias of XIS solution close to zero
   is k=0.78, somewhat smaller than that from solution XIB based on data 
   from the VLBA network. Moreover, estimates of declinations from XIS solution 
   require a different de-bias correction than declinations from XIB solution. 
}

\begin{figure}
   \includegraphics[width=0.47\textwidth]{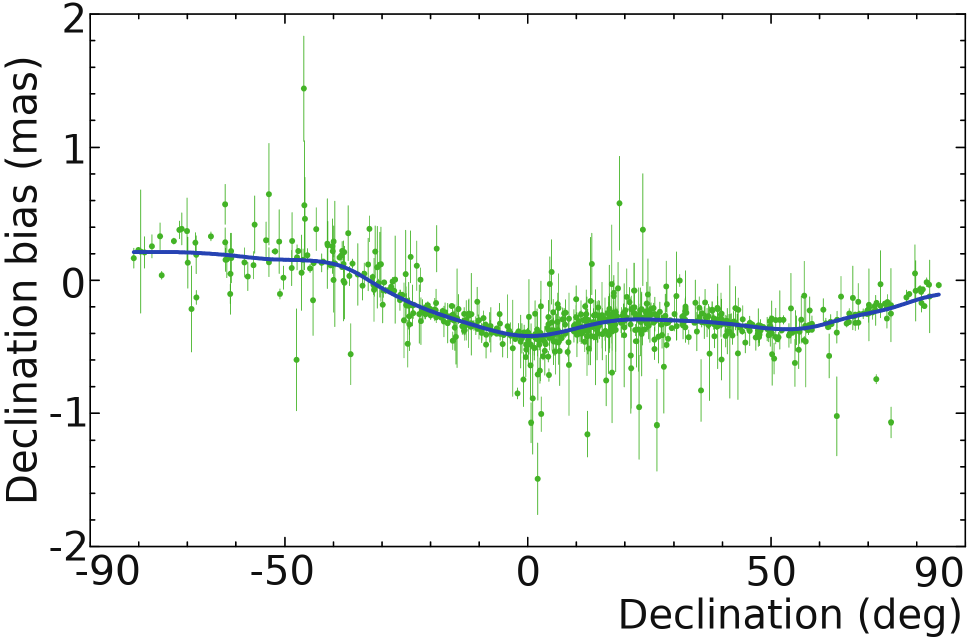}
   \caption{Differences in declinations from the X-band solution
            XIR with respect to the dual reference solution RIR
            using data at the R1/R4 VLBI, similar to 
            Figure~\ref{f:x_bias_1000}. Default TEC GNSS mapping
            function ($\Delta H=0, \alpha=1.0, k=1.00$)
            was used that comparison.
           }
   \label{f:x_r1r4_bias_1000}
\end{figure}

\Note{
   Figure~\ref{f:x_r1r4_bias_1000} shows differences in declinations from 
   position estimates of 638 sources with formal uncertainties $< 0.5$~mas
   from XIR solution with respect to the reference solution RIR based on
   data from R1/R4 networks. Comparing solutions made with different 
   mapping function parameters $k$, I found that the mean declination 
   bias vanishes when $k=0.75$. The shape of the bias and its magnitude is 
   noticeably different than the bias from declinations derived using
   group delays from observations at the VLBA network 
   (see Figure~\ref{f:x_bias_1000}).
}

\Note{
   The magnitude of the peak-to-peak declination bias from the solution that 
   uses data from the R1/R4 network is 0.62~mas, while the magnitude
   of the declination bias from the solution that uses data from the VLBA
   network is 0.85~mas. The regional VLBA network stretches over 
   $90^\circ$ in longitudes, but only within $[18^\circ, 48^\circ]$ in 
   latitude, while the global R1/R4 network stretches over all longitudes 
   and over $[-43^\circ, +79^\circ]$ in latitude. Observations of 
   Southern Hemisphere sources with VLBA are possible in the southern 
   sectors only, and the further south a source, the lower elevations it 
   can be observed. At the same time, a given source can be observed at 
   different azimuths and elevations at a global network.
}

\begin{figure}[h]
   \par\bigskip\par
   \caption{Similar as Figure~\ref{f:iono_mod_stat} but derived from
            analysis of observations at R1/R4 VLBI network. Statistics
            computed for one year intervals.}
   \includegraphics[width=0.47\textwidth]{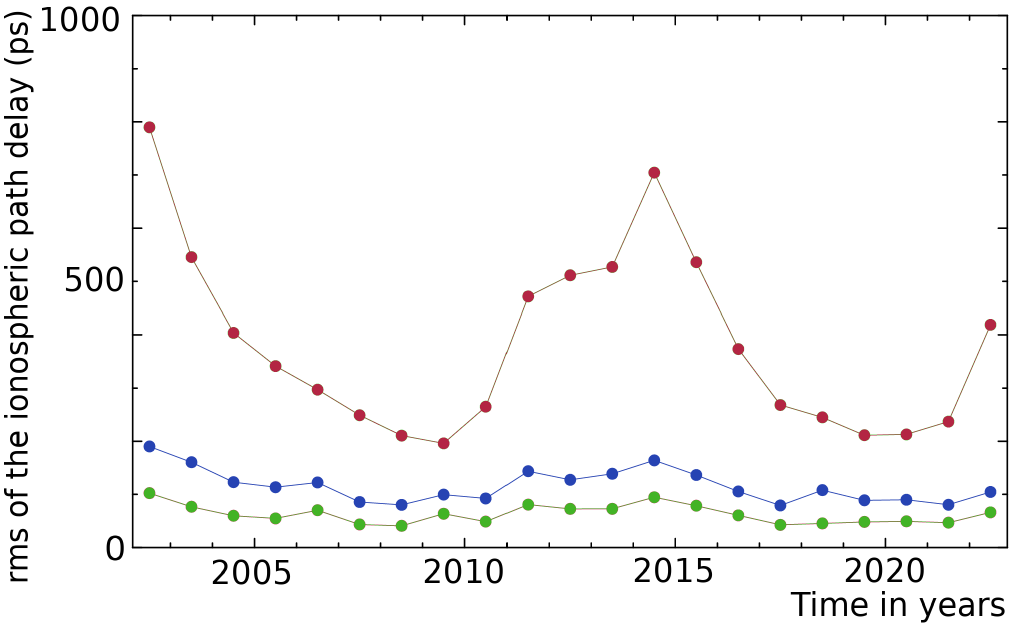}
   \label{f:iono_r1r4_mod_stat}
\end{figure}

\Note{
   Figure~\ref{f:iono_r1r4_mod_stat} shows the rms of the mean residual 
ionospheric contribution at R1/R4 baselines. Comparing it with a similar
plot for the VLBA network (Figure~\ref{f:iono_mod_stat}), we see that
the wrms of the ionospheric contribution at the VLBA network is 
a factor of 2--3 lower. This can be explained by both differences in
total ionospheric path delay sensed by stations at two networks, and by
the amount of the ionospheric path delay that is absorbed in estimates
of parameters, such as source declination, causing biases. If a source
is observed only at a given azimuth and elevation, most of the ionospheric
path delay will be absorbed by estimated parameters reducing residuals.
The wider spread in azimuths and elevations a given source is observed, 
the smaller share of the ionospheric path delay that will be absorbed by 
estimated parameters and propagate to residuals.
}

\Note{
   We can estimate the scaling factor directly from single-band VLBI
data using least squares. I ran twelve solutions using data from the 
VLBA, southern, and R1/R4 networks separately and using data from all 
three networks altogether and applying three different flavors of the 
ionospheric mapping function, $\Delta H$=0, $\alpha=1.0$; $\Delta H$=56.7, 
$\alpha=0.9782$; and $\Delta H$=150,  $\alpha=0.9782$. Table~\ref{t:k_est}
displays the estimates. The differences between the estimates of the 
scaling factors derived from observations at the R1/R4 and Southern 
Hemisphere networks are statistically insignificant, while the scaling 
factor estimates from the VLBA network are systematically lower, and 
these differences with respect to the R1/R4 network are statistically 
significant. It is also remarkable that scaling factors that minimize 
postfit residuals differ from those that eliminate the mean declination 
bias. This implies that scaling does not eliminate systematic errors 
due to deficiency of GNSS TEC maps but only alleviates their impact.
}

\begin{table}[h]
   \caption{
            \note{
            Estimates of scaling factors from twelve solutions at 
            the Southern Hemisphere, VLBA, and R1R4 network, as well
            as all network combined. Three different ionospheric
            mapping functions were used: 
            1: $\Delta H=0,    \alpha=1.0,    k=1.00$,
            2: $\Delta H=56.7, \alpha=0.9782, k=1.00$,
            3: $\Delta H=150,  \alpha=0.9782, k=1.00$.
           }
           }
   \noindent\hspace{-0.18\textwidth}
   \begin{tabular}{l l@{\;}c@{\;}l l@{\;}c@{\;}l l@{\;}c@{\;}l}                    
        \hline
        Network & \nnnnnnnnntab{c}{Ionospheric mapping functions}                       \\
                & \nnntab{c}{(1)}       & \nnntab{c}{(2)}       & \nnntab{c}{(3)}       \\
        \hline
        VLBA    & 0.750 & $\pm$ & 0.002 & 0.801 & $\pm$ & 0.002 & 0.835 & $\pm$ & 0.002 \\
        R1R4    & 0.789 & $\pm$ & 0.001 & 0.844 & $\pm$ & 0.001 & 0.882 & $\pm$ & 0.001 \\
        South   & 0.79  & $\pm$ & 0.02  & 0.84  & $\pm$ & 0.02  & 0.87  & $\pm$ & 0.02  \\
        all     & 0.780 & $\pm$ & 0.001 & 0.834 & $\pm$ & 0.001 & 0.872 & $\pm$ & 0.001 \\
        \hline
   \end{tabular}
   \label{t:k_est}
\end{table}

\subsection{A combined use of dual-band and single-band data}

  \Note{Negligible remaining} biases and the realistic assessment of errors 
due to the residual ionospheric contribution allows us to use a mixture of 
dual-band and single-band group delays in a single least square solution. 
Source positions derived from single-band data are in general less precise 
than those derived from dual-band data because an additional factor affects 
the uncertainties of  group delays. These additional errors can be computed 
and accounted for in deriving weights of observables. An increase in 
uncertainties of group delay observables propagates to uncertainties of 
source positions. 

  Combined analysis of a heterogeneous datasets provides a significant 
advantage because all the data are used. We can fuse dual-band and single-band
data in one dataset and use it for estimation of source positions.
A fused dataset consists of observables of three types:

\begin{enumerate} 
  \item Dual-band ionosphere-free linear combinations of group delays.

  \item Single-band delays of dual-band experiments detected at one band only.
        Data reduction accounts for the ionospheric path delay computed
        from the sum of GNSS TEC and TEC bias adjustments. The uncertainty of 
        the bias adjustment is added in quadrature to the group delay 
        uncertainty for reciprocal weights of such observables.

  \item Group delays from single-band experiments. Data reduction accounts
        for the ionospheric path delay computed from GNSS TEC and applies
        the de-bias correction. The uncertainty of the ionospheric model 
        computed from the regression model is added in quadrature to the 
        group delay uncertainty used for computation of weights of such 
        observables.
\end{enumerate}

   Solving for source coordinates using a fused dataset provides estimates 
with minimum uncertainties and minimum correlations \note{between right 
ascensions and declinations} since it uses all available data provided
the frequency-dependent position biases due to source structure are 
negligible. This condition may be violated for some strong sources with 
significant structure. Analysis of position offsets between single band and 
dual-band solutions will help to identify such sources 
\citep[see, for example,][]{r:vgaps}, but these cases are encountered 
infrequently.

\subsection{Future work}

  An improvement in modeling of ionospheric path delay when TEC biases 
are adjusted is quite impressive. However, this requires using additional 
information about the state of the ionosphere. Specifically, VLBI 
ionospheric path delays from dual-band data were used for computation of 
these biases. This information is not available in processing of 
single-band observations. \citet{r:ros00} considered the use of dual-band 
GPS observations from collocated receivers for analysis of radio astronomy 
observations. They have shown that the GPS determination of TEC from ground 
receivers alone without the use of GNSS TEC maps can be successfully 
applied to the astrometric analysis of VLBI observations.

  There are on-going efforts to install advanced GNSS receivers within a hundred 
meters of each VLBA antenna. The use of geometry-free pseudo-ranges at 1.2 and 
1.5~GHz in a similar way as I used VLBI ionosphere-free group delays for 
adjustments of TEC biases promises a similar level of improvement.

\section{Conclusions}

  Single-band group delays are affected by the contribution of the ionosphere.
This contribution is noticeable at frequencies below 30~GHz and becomes 
a dominating error source at frequencies below 5--8~GHz. Compared with
astrometric solutions based on the use of the ionosphere-free linear 
combinations of dual-band observables, \note{source position estimates derived 
from a single-band solution are} affected by additional random and 
systematic errors caused by mismodeling the contribution of the ionosphere 
to path delay. I explored two approaches to modeling the ionospheric path 
delay using GNSS TEC maps and assessed residual ionospheric errors.

  The findings can be summarized as follows:

\begin{enumerate} 
    \item In a case when an experiment was recorded at two widely separated 
          bands, but a fraction of sources were detected at one band only, 
          estimation of the TEC bias in a form of an expansion over the B-spline 
          basis using dual-band data and then  applying that bias to 
          GNSS TEC maps provides an unbiased estimates of sources positions. 
          The stochastic model that describes residual errors of TEC bias 
          adjustment predicts an increase of positions errors with an accuracy 
          of 15\%. No remaining systematic errors were found. This approach 
          provides source positions with the lowest uncertainties with respect 
         to other approaches.

    \item In a case of single-band observations, path delays are  computed 
          using GNSS TEC maps. The thin spherical shell model of the 
          ionosphere with the constant height 450~km above the mean Earth's 
          radius causes a strong systematic bias in declinations that reaches 
          1~mas at 8~GHz an 12~mas at 2.3~GHz. This bias can be virtually 
          eliminated when a)~the modified ionospheric mapping function  with 
          parameters $\Delta H=$56.7~{\rm km},\, $\alpha=0.9782,$\, $k=0.85$ is 
          used; and b)~the empirical de-bias correction is applied.

    \item Ionospheric errors are mainly multiplicative. 
          
    \item \Note{I determined the scaling factor of TEC from GNSS maps by
          processing from 11,457,749 VLBI observations at different
          networks for 25 years: 0.834 $\pm$ 0.001. This estimate}
          is in a good agreement with a totally 
          independent comparison of vertical TEC determined from satellite 
          altimetry against GNSS TEC maps. Since VLBI is sensitive to a delay
          incurred in the total ionosphere, interpretation of a scaling factor 
          of TEC from altimetry and radio occultation observations as 
          the contribution of upper layers of the ionosphere at altitudes above 
          Jason orbit, i.e. 1,350~km, suggested by \citet{r:liu18} and 
          \Note{\citet{r:det22,r:det23} is not consistent with presented 
          results. Although the plasmosphere above 1,350~km may contribute 
          to discrepancies between Jason and GNSS TEC, systematic errors
          in GNSS TEC due to the adopted model of vertical distribution 
          of electron density in a form of a thin shell at height 450 or 
          506.7~km is another major factor. The contribution of the 
          plasmosphere alone cannot explain the scaling factor.
          }

    \item I have found that the scaling factor of GNSS TEC maps that provides 
          zero mean declination bias depends on the used ionospheric 
          mapping function. Therefore, I surmise that the established 
          deficiency of GNSS TEC maps is caused by over-simplification of the
          ionospheric mapping function used for their derivation that considers
          the ionosphere as a thin spherical shell. The electron contents 
          in the real ionosphere varies not only in latitude and longitude, 
          but also in height. Diurnal variations of the effective ionosphere 
          height at a level of 100~km, i.e. over 20\%, are large enough to cause
          errors of the magnitude that was found in analysis of VLBI data.

    \item The impact of the ionosphere on path delay depends on the Solar 
          cycle. Modeling ionospheric path with GNSS TEC maps reduces the 
          residuals at 8~GHz by a factor of 2 during the Solar maximum and
          only by 10\% during the solar minimum. Estimation of the TEC bias
          reduces ionospheric errors further by a factor of 2 regardless of the 
          phase of the Solar maximum.
          
    \item The impact of the ionosphere on source position errors can be
          modeled with an accuracy of 15\%. It remains noticeable even at 
          frequencies as high as 22--24~GHz (K-band). In particular, the 
          ionospheric errors even after applying data reduction based on
          GNSS TEC maps with the modified mapping function and the de-bias 
          correction exceed 0.1~mas. Declination errors of \note{Southern
          Hemisphere sources observed with VLBA} are in a range of 0.1 to 
          0.3~mas. An assertion that K-band astrometry is able to provide 
          results more precise than 0.1~mas is not true at the current state 
          of our ability to model ionospheric path delay. Considering on-going 
          efforts to install advanced GNSS receivers in the close vicinity 
          of VLBA stations and other radio telescopes, the situation may 
          change in the future.
\end{enumerate}

\par\vspace{-1ex}\par

  This study lays the foundation of the single-band absolute astrometry. 
Dual-band astrometric observations still provide the best accuracy.
The use of singe-band data with the procedure of data reduction and 
weighting described above allows us to get unbiased positions with
{\it known} added errors.

\par\vspace{-2ex}\par
\section{Acknowledgments}
\begin{acknowledgments}

   This work was done with datasets RDV, RV, CN, UF001, UG003, BL122, BL166, 
BP133, BP138, GC073, BC204, BG219, and V17 collected with VLBA instrument of 
the NRAO and available at  \web{https://archive.nrao.edu/archive}. The NRAO 
is a facility of the National Science Foundation operated under cooperative 
agreement by Associated Universities, Inc. \note{The author acknowledges use 
of the VLBA under the USNO's time allocation for some datasets.} This work made 
use of the Swinburne University of Technology software correlator, developed 
as part of the Australian Major National Research Facilities Programme and 
operated under license.

   It is my pleasure to thank Yuri Y.~Kovalev, Urs Hugentobler,  Robert 
Heinkelmann, Minghui Xu, \note{and Denise Dettmering} for discussions 
of presented results.
\end{acknowledgments}

\facility{VLBA}
\software{PIMA}

\bibliographystyle{aasjournal}
\bibliography{sba}

\appendix

\begin{figure}[h]
   \par\bigskip\par
   \caption{The empirical declination bias corrections $D(\delta)$ computed 
            from the coefficients presented in Table~\ref{t:d_expansion}
            for three VLBI networks: the Southern Hemisphere network 
            marked with $S$ (red), R1/R4 marked with $R$ (blue), and
            VLBA marked with $V$ (green).
           }
   \includegraphics[width=0.47\textwidth]{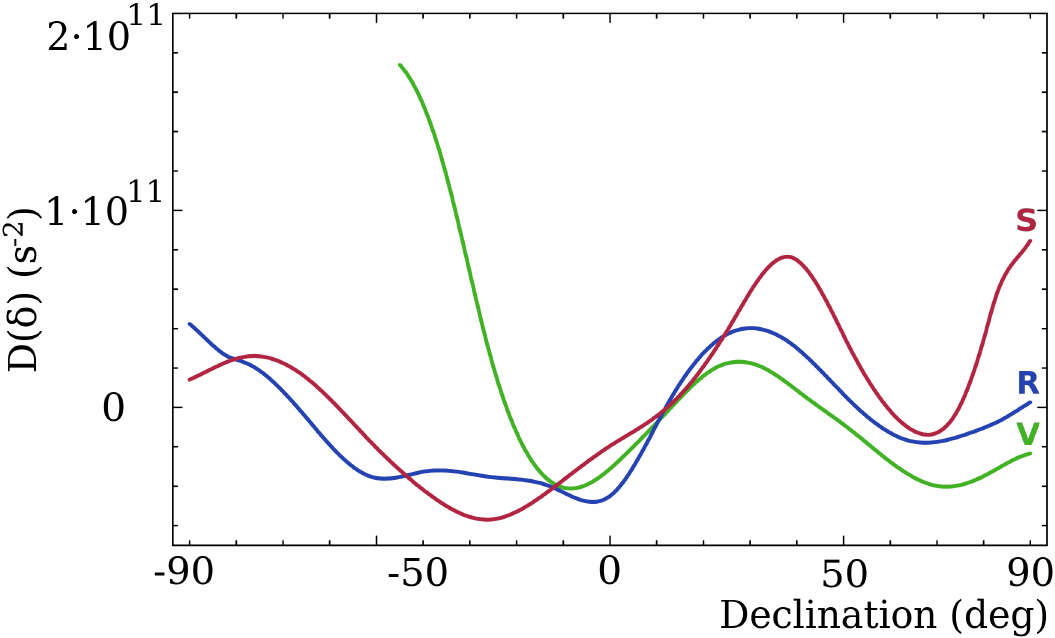}
   \label{f:da}
\end{figure}

\Note{
   Coefficients of the expansion of the empirical correction of declination 
bias into the B-spline basis of the 3d degree due to deficiency of TEC GNSS 
maps for three networks are given in Table~\ref{t:d_expansion}.
Plots of the empirical corrections $D(\delta)$ computed using these 
coefficients are shown in Figure~\ref{f:da}.
}

\begin{table}[h]
   \caption{Coefficients of the expansion of declination 
            correction for three network into the B-spline 
            basis of the 3rd degree. 
           }
   \noindent\hspace{-2.25em}
   \parbox[t]{0.33\textwidth}{
      \begin{tabular}{rrl}
             \hline
              \nnntab{c}{VLBA network}                              \\
              \nnntab{c}{$\Delta H=56.7~{\rm km},\, \alpha=0.9782,\, k=0.85$} \\
              knot &  \ntab{c}{$\delta$} & \ntab{c}{$D_k(\delta)$}  \\
                   &  \ntab{c}{rad}      & \ntab{c}{$s^{-2}$}       \\
             \hline
                -2 &  -0.78539 & \hp $  1.7254  \cdot 10^{+11} $    \\
                -1 &  -0.78539 & \hp $  1.6909  \cdot 10^{+11} $    \\
                 0 &  -0.78539 & \hp $  1.1094  \cdot 10^{+11} $    \\
                 1 &  -0.78539 &     $ -1.2356  \cdot 10^{+10} $    \\
                 2 &  -0.58904 &     $ -4.7934  \cdot 10^{+10} $    \\
                 3 &  -0.39269 &     $ -3.3498  \cdot 10^{+10} $    \\
                 4 &  -0.19634 &     $ -4.9275  \cdot 10^{+09} $    \\
                 5 &   0.00000 & \hp $  2.5485  \cdot 10^{+10} $    \\
                 6 &   0.19634 & \hp $  2.2408  \cdot 10^{+10} $    \\
                 7 &   0.39269 &     $ -1.2769  \cdot 10^{+09} $    \\
                 8 &   0.58904 &     $ -1.9056  \cdot 10^{+10} $    \\
                 9 &   0.78539 &     $ -4.3368  \cdot 10^{+10} $    \\
                10 &   0.98174 &     $ -3.8566  \cdot 10^{+10} $    \\
                11 &   1.17809 &     $ -2.5999  \cdot 10^{+10} $    \\
                12 &   1.37444 &     $ -2.3141  \cdot 10^{+10} $    \\
                13 &   1.57079 &     $ -2.3141  \cdot 10^{+10} $    \\
                   &           &                                    \\
                   &           &                                    \\
             \hline 
      \end{tabular}
   }
   \parbox[t]{0.33\textwidth}{
      \begin{tabular}{rrl}
             \hline
              \nnntab{c}{Southern network}                         \\
              \nnntab{c}{$\Delta H=56.7~{\rm km},\, \alpha=0.9782,\, k=0.78$} \\
              knot &  \ntab{c}{$\delta$} & \ntab{c}{$D_k(\delta)$} \\
                   &  \ntab{c}{rad}      & \ntab{c}{$s^{-2}$}      \\
             \hline
                -2 &  -1.57079 & \hp $  1.4031 \cdot 10^{+10} $    \\
                -1 &  -1.57079 & \hp $  1.8702 \cdot 10^{+10} $    \\
                 0 &  -1.57079 & \hp $  3.1684 \cdot 10^{+10} $    \\
                 1 &  -1.57079 & \hp $  1.7161 \cdot 10^{+10} $    \\
                 2 &  -1.34639 &     $ -1.8306 \cdot 10^{+10} $    \\
                 3 &  -1.12199 &     $ -4.6113 \cdot 10^{+10} $    \\
                 4 &  -0.89759 &     $ -6.3192 \cdot 10^{+10} $    \\
                 5 &  -0.67319 &     $ -4.2758 \cdot 10^{+10} $    \\
                 6 &  -0.44879 &     $ -1.7912 \cdot 10^{+10} $    \\
                 7 &  -0.22439 &     $ -3.2968 \cdot 10^{+09} $    \\
                 8 &   0.00000 & \hp $  3.8267 \cdot 10^{+10} $    \\
                 9 &   0.22439 & \hp $  9.8592 \cdot 10^{+10} $    \\
                10 &   0.44879 & \hp $  2.5145 \cdot 10^{+10} $    \\
                11 &   0.67319 &     $ -1.9324 \cdot 10^{+10} $    \\
                12 &   0.89759 &     $ -1.3308 \cdot 10^{+10} $    \\
                13 &   1.12199 & \hp $  9.5708 \cdot 10^{+10} $    \\
                14 &   1.34639 & \hp $  8.4468 \cdot 10^{+10} $    \\
                15 &   1.57079 & \hp $  8.4468 \cdot 10^{+10} $    \\
             \hline 
      \end{tabular}
   }
   \parbox[t]{0.33\textwidth}{
      \begin{tabular}{rrl}
             \hline
              \nnntab{c}{R1/R4 network}                            \\
              \nnntab{c}{$\Delta H=56.7~{\rm km},\, \alpha=0.9782,\, k=0.75$} \\
              knot &  \ntab{c}{$\delta$} & \ntab{c}{$D_k(\delta)$} \\
                   &  \ntab{c}{rad}      & \ntab{c}{$s^{-2}$}      \\
             \hline
                -2 &  -1.57079 & \hp $  4.2459 \cdot 10^{+10} $    \\
                -1 &  -1.57079 & \hp $  2.4792 \cdot 10^{+10} $    \\
                 0 &  -1.57079 & \hp $  2.8538 \cdot 10^{+10} $    \\
                 1 &  -1.57079 &     $ -7.0868 \cdot 10^{+09} $    \\
                 2 &  -1.34639 &     $ -4.3644 \cdot 10^{+10} $    \\
                 3 &  -1.12199 &     $ -2.8135 \cdot 10^{+10} $    \\
                 4 &  -0.89759 &     $ -3.7050 \cdot 10^{+10} $    \\
                 5 &  -0.67319 &     $ -3.5487 \cdot 10^{+10} $    \\
                 6 &  -0.44879 &     $ -6.1013 \cdot 10^{+10} $    \\
                 7 &  -0.22439 & \hp $  9.6720 \cdot 10^{+09} $    \\
                 8 &   0.00000 & \hp $  4.5049 \cdot 10^{+10} $    \\
                 9 &   0.22439 & \hp $  3.7563 \cdot 10^{+10} $    \\
                10 &   0.44879 & \hp $  1.0745 \cdot 10^{+09} $    \\
                11 &   0.67319 &     $ -2.2499 \cdot 10^{+10} $    \\
                12 &   0.89759 &     $ -1.3549 \cdot 10^{+10} $    \\
                13 &   1.12199 &     $ -5.1436 \cdot 10^{+09} $    \\
                14 &   1.34639 & \hp $  2.6590 \cdot 10^{+09} $    \\
                15 &   1.57079 & \hp $  2.6590 \cdot 10^{+09} $    \\
             \hline 
      \end{tabular}
   }
  \label{t:d_expansion}
\end{table}

\end{document}